\documentclass[final,5p,times,twocolumn]{elsarticle}

\usepackage{amssymb}
\usepackage{amsthm}
\usepackage{amsmath}
\usepackage{algorithmic}
\usepackage{color}
\usepackage{textcomp}
\newcommand\rfrac[2]{{#1}/{#2}}

\newcommand\D{{\bf d}}
\newcommand\E{{\bf e}}

\newcommand\G{{\bf g}}
\newcommand\M{{\bf m}}
\newcommand\N{{\bf n}}

\newcommand\F{{\bf f}}
\newcommand\T{{\bf t}}
\newcommand\U{{\bf u}}

\newcommand\X{{\bf x}}

\renewcommand\AA{{\bf A}}

\newcommand\DD{{\bf D}}

\newcommand\LL{{\bf L}}

\newcommand\dert{\partial_t}

\newcommand\grad{\nabla}

\newcommand\eps{\epsilon}

\newcommand\be{\begin{equation}}
\newcommand\nd{\end{equation}}
\newcommand\bed{\begin{displaymath}}
\newcommand\ndd{\end{displaymath}}

\newcommand\ba{\begin{array}}
\newcommand\ea{\end{array}}
\newcommand\bea{\begin{eqnarray}}
\newcommand\nda{\end{eqnarray}}

\newcommand\Order{{\cal O}}
\newcommand\cijk{C_{i,j,k}}

\def\vec#1{\ensuremath{\mathchoice
                       {\mbox{\boldmath$\displaystyle#1$}}
		       {\mbox{\boldmath$\textstyle#1$}}
		       {\mbox{\boldmath$\scriptstyle#1$}}
		       {\mbox{\boldmath$\scriptscriptstyle#1$}}}}%

\def\ba{\vec{a}}

\def\be{\vec{e}}

\def\bi{\vec{i}}

\def\BCancel{\kern-2.em\Bigg\backslash\kern+0.7em}
\def\Bcancel{\kern-1.em\Bigg\backslash\kern-0.2em}
\def\bcancel{\kern-1.em\bigg\backslash\kern-0.0em}
\def\cancel{\kern-0.8em\big\backslash\kern+0.2em}

\def\grad{\hbox{grad}\ }

\font\bigestssbx=cmssbx10     scaled \magstep4    
\def\assemble{\mathop{\lower 5.0pt \hbox{\bigestssbx A}}}
\def\Passemble{\mathop{\lower 5.0pt \hbox{\bigestssbx P}}}

\def\dP{\D\,\raise1pt\hbox{$\sqr{7}{4}$}}

\def\be{\begin{equation}}
\def\ee{\end{equation}}
\def\ba{\begin{array}}
\def\ea{\end{array}}
\def\bfr{\begin{frame}}
\def\efr{\end{frame}}
\def\bi{\begin{itemize}}
\def\ei{\end{itemize}}

\newcommand\ubar{{\bar u}}

\def\ubar{{\bf u}}

\def\grad{{\nabla}}

\newcommand\La{{\rm La}\,}

\newcommand\infinity\infty
\newcommand\pariss{{\sc Paris} }
\newcommand\paris{{\sc Paris}}

\newcommand\gerris{{\sc Gerris}}
\newcommand\gerriss{{\sc Gerris} }
\newcommand\basilisk{{\sc Basilisk}}
\newcommand\basilisks{{\sc Basilisk} }
\newcommand\surfers{{\sc Surfer} }
\newcommand\surfer{{\sc Surfer}}
\newcommand\gretars{{\sc Ftc3D} }
\newcommand\gretar{{\sc Ftc3D}}
\newcommand\Heaviside{\chi}
\newcommand\markerfunction{I}

\newcounter{bla}

\journal{Computer Physics Communications}

\newcommand {\red}[1]{{#1}}

\newcommand {\green}[1]{{\color{green}{#1}}}

\begin{document}

\begin{frontmatter}



\title{{PA}rallel, Robust, Interface Simulator (PARIS) }

\author[b]{W. Aniszewski}
\author[b]{T. Arrufat}
\author[f]{M. Crialesi-Esposito}
\author[nd]{S. Dabiri}
\author[b]{D. Fuster}
\author[b,bu]{Y. Ling}
\author[label1]{J. Lu}
\author[b,d]{L. Malan}
\author[b]{S. Pal}
\author[e]{R. Scardovelli}
\author[label1]{G. Tryggvason}
\author[g]{P.~Yecko}
\author[b]{S. Zaleski}
\address[b]{Sorbonne Universit\'e \& CNRS, UMR 7190, \\
Institut Jean Le Rond d'Alembert, F-75005, Paris, France}
\address[f]{CMT-Motores T\'ermicos, Universitat Polit\'ecnica de Val\'encia, 
Camino de Vera, s/n, Edificio 6D, Valencia, Spain}
\address[nd]{Mechanical Engineering, Purdue University, West Lafayette, IN, USA}
\address[bu]{Mechanical Engineering, Baylor University, Waco, TX 76706, USA}
\address[d]{Mechanical Engineering, University of Cape Town, South Africa}
\address[label1]{Mechanical Engineering, Johns Hopkins University, Baltimore, MD, USA}
\address[e]{DIN - Lab. di Montecuccolino, Universit\`a di Bologna, I-40136 Bologna, Italy}
\address[g]{Cooper Union, New York City, USA}

\begin{abstract}

\pariss ({PA}rallel, Robust, Interface Simulator) is a finite volume code for simulations 
of immiscible \red {multifluid} or multiphase flows.  It is based on the ``one-fluid'' 
formulation of the Navier-Stokes equations  where different fluids are treated as one 
material with variable properties, and surface tension is added as a singular interface force. 
The fluid equations are solved on a regular structured staggered grid using \red {an  
explicit projection method with a first-order or second-order time integration scheme}. 
The interface separating the different fluids is tracked \red{by} 
a Front-Tracking (FT) method, where the interface is represented by connected marker points, 
or by a Volume-of-Fluid (VOF) method, where the marker function is advected directly on the 
fixed grid. \pariss is written in Fortran95/2002 and parallelized using MPI and domain 
decomposition. It is based on several earlier FT or VOF codes such as \gretar, \surfers or 
\gerris. These codes and similar ones, as well as \paris, have been used to simulate a wide 
range of multifluid and multiphase flows.

\end{abstract}

\end{frontmatter}

\newcommand\LLL{{\cal L}}



{\bf PROGRAM SUMMARY}

\begin{small}
\noindent
{\em Program Title:}   PArallel Robust Interface Simulator --- \paris \\
{\em Licensing provisions:} GPLv3.                                   \\
{\em Programming language:} Fortran95/2002. Parallelized using MPI and domain decomposition.                                  \\
{\em Nature of problem:}\\
\pariss is a free code, or software, for computational fluid dynamics (CFD)
of multiphase flows \red {(or computational multiphase fluid dynamics
(CMFD)), specialized in the numerical simulation of interfacial fluid flows, involving}
droplets, bubbles and waves, as described 
in the book by Tryggvason, Scardovelli and Zaleski [1]. 
It solves the Euler or Navier-Stokes
equations in the one-fluid formulation of two-phase flow, \red {including
a surface tension force}.
It computes complex flows such as fast atomizing jets or droplets, 
expanding cavitation bubble clusters and multiphase flow through porous media.
 \\
{\em Solution method:}\\
The code mostly implements the methods described
in the book by Tryggvason, Scardovelli and Zaleski [1]. 
Time stepping is performed using \red {a first-order or a second-order} in time 
predictor-corrector method
with an explicit projection \red {step} for the pressure. 
Spatial \red {discretization} is by finite volumes on a regular cuboid grid. 
Interface tracking is performed 
with the Front-Tracking (FT) method or the Volume-of-Fluid (VOF) method. 
In the VOF version \pariss uses either the Lagrangian-Explicit (LE) advection 
method or the exactly mass-conserving method of Weymouth and Yue [2]. 
The normal computation is performed using the Mixed-Youngs-Centered (MYC) 
scheme.
\red {A mass-momentum advection method has been also implemented that is 
consistent with the VOF advection} [3]. 
Curvature is computed with the Height Function (HF) method. This is 
combined with the balanced Continuous Surface 
Force (CSF) method to compute surface tension forces. 
\red 
{If the dynamics of a phase can be neglected, \pariss can also run as a 
free-surface code by specifying a homogeneous pressure, at most varying 
with time, in the neglected phase. In the case of atomizing jets, an algorithm
has been implemented in \pariss that can detect isolated droplets,
advects them as Lagrangian point-particles and possibly merge them
again with the main stream.} \\
{\em Additional comments:}\\
\pariss is extended from or inspired by the following codes:
\begin{itemize}
\item \gretar: Front Tracking code for 3D simulations 
by Gretar Tryggvason and Sadegh Dabiri
\item \surfer: VOF code for 3D simulations by Stephane Zaleski, Jie Li, 
Ruben Scardovelli and others
\item \gerris: multiphase flow solver with Adaptive Mesh Refinement (AMR) 
by Stephane Popinet
\end{itemize}

\end{small}


\newcommand\division[1]{\subsection{#1}}
\newcommand\subdivision[1]{\subsubsection{#1}}
\newcommand\onlybook[1]{{}}
\newcommand\opus{article}
\newcommand\reduit[1]{#1}

\section{Introduction} 
\label{introduction}
Computations of the unsteady motion of \red {multifluid} flows, where two or
more immiscible fluids or thermodynamic phases 
flow while separated by sharp interfaces,
date back to the earliest days of computational fluid dynamics
(see \cite{scardovelli99,Tryggvason11} for reviews). 
However, early simulations were
restricted to relatively small and idealized problems. As computer
power has continued to grow, it has been  increasingly possible to
conduct Direct Numerical Simulations (DNS), defined as fully resolved
and verified simulations of a validated system of equations that
include non-trivial length and time scales. 

\red {A few authors of \pariss \citep{parissimulator} have been involved in 
the development of {\em free codes} for DNS of two-phase flows,
such as \surfers \cite{Lafaurie94}, 
\gerriss \citep{gerris-web}, dating back a couple of decades,
and more recent ones such as \basilisks \citep{basilisk}. The new project
\pariss is a joint effort with the aim to illustrate most of the methods 
described in \cite{Tryggvason11} 
(the latter book will be denoted ``TSZ'' in what follows to simplify citations and 
we will often refer to ``TSZ'' for developments about the numerical methods)
and it relies heavily on the previous codes \gretars and \surfer.
The software can run independently either with the Front-Tracking method or 
the VOF method. The mass-momentum consistent advection method, the free-surface
option and the Lagrangian point-particles (LPP) algorithm have been specifically
designed for \pariss and were not present in the previous codes.} 

\red {Furthermore, a software platform such as Paris can make it possible to 
explore in the near future the development of combined models based on both
Front-Tracking and Volume-of-Fluid methods.}

While \red {DNS of multiphase flows} are becoming increasingly common, 
most research groups need to devote \red {a considerable amount of time} 
to code development. 
For new groups, the need to develop a suitable simulation tool can be a 
significant barrier to entry. \pariss is a free code that is intended to be
relatively simple to use and modify \red {even by beginner users}, yet it has 
sufficient capabilities to allow state-of-the-art studies of typical multiphase 
flow problems.

\section{Navier--Stokes equations with interfaces} 
\label{nse}
\subsection{Basic equations}

In \red {the} three-dimensional space, the locus of the interface is a smooth 
surface $S$ which separates the two fluid phases.
Accordingly, we \red {assume} that the interface is an object of zero thickness.
This latter assumption constitutes the ``sharp interface'' approximation. 
In this approximation, the phases are implicitly located by a Heaviside function 
$\Heaviside(\X,t)$ defined such that fluid 1 corresponds to $\Heaviside=1$ and 
fluid 2 to $\Heaviside=0$. 
Viscosity and density $\mu$ and $\rho$ are space and time dependent and given by
\be
\mu = \mu_1 \Heaviside + \mu_2 (1-\Heaviside)\,, \qquad \rho = \rho_1 \Heaviside + 
\rho_2 (1 - \Heaviside) \,. 
\label{muH}
\nd
In the case with no phase change, mass conservation implies that the interface 
advances at the speed of the flow, that is
\be
V_S=\U(\X,t) \cdot \N 
\label{vinterf}
\nd
where $\U(\X,t)$ is the local fluid velocity and $\N$ a unit normal vector 
perpendicular to the interface. Equivalently, this condition on
the interface motion can be expressed, in weak form, as
\be
\dert \Heaviside + \U \cdot \grad \Heaviside = 0 \,, 
\label{interfadv}
\nd
which expresses the fact that the singularity of $\Heaviside$, located on $S$, 
moves at velocity $V_S=\U\cdot \N$. We refer the reader to the literature, 
in particular TSZ, for additional developments on interface geometry. 

For incompressible flows, which we will consider in what follows, we have
\be
\nabla \cdot \U = 0 \,.
\label{divu}
\nd
The Navier--Stokes equations for incompressible, Newtonian flow with surface tension 
may conveniently be written in a conservative form, expressing the momentum balance, 
or in a non conservative, Lagrangian form. The first form, using operators for 
notational simplicity, is 
\be
 \dert (\rho \U) = \LLL_1(\rho,\U) - \grad p 
 \label{nscons}
\nd
where 
$
\LLL_1 =  \LLL_{\rm cons} + \LLL_{\rm diff} +   \LLL_{\rm cap} + \LLL_{\rm ext}
$
so that the operator $\LLL_1$ is the sum of a conservative momentum transport 
term and diffusive, capillary and external force terms. The first two terms are 
\be
\LLL_{\rm cons} = -\grad \cdot ( \rho \U  \U )\,, \qquad \LLL_{\rm diff} =  
\grad \cdot \DD \,,
\label{diff1}
\nd
where $\DD$ is the stress tensor whose expression for incompressible flow is
\be
\DD = 2 \,\mu \, \red {
{\bf S} \,, \quad {\bf S} = \tfrac{1}{2}  \big( \nabla\U + (\nabla\U)^T \big) \,,}
\label{diff2}
\nd
and $\mu$ is computed from $\Heaviside$ using (\ref{muH}). 
The capillary term is
\be
\LLL_{\rm cap} =  {\sigma \,\kappa \,\delta_S \,\N} + \nabla_S \sigma \,\delta_S\,,  
\qquad \kappa = 1/R_1 + 1/R_2 \,, 
\label{lllcap} 
\nd
where $\sigma$ is the surface tension coefficient, \red {$\nabla_S \sigma$ its surface 
gradient}, $\N$ the unit normal perpendicular to the interface,
$\kappa$ the sum of principal curvatures and $\delta_S$ 
a Dirac distribution concentrated on the interface. \red {Other equations are
required to specify the dependence of $\sigma$ from other physical quantities, 
such as temperature, the presence of surfactants or electro-magnetic fields,
therefore in the tests section we will assume a constant $\sigma$}.  
Finally $\LLL_{\rm ext}$ represents external forces. When the external force is gravity
$\LLL_{\rm ext} = \rho \,\G$.

 The second, non conservative form, is 
\be
\dert \U = \LLL_2(\rho,\U) - \frac 1 \rho \grad p 
\label{nsnoncons}
\nd
where 
$\LLL_2 =  \LLL_{\rm adv} +  ( \LLL_{\rm diff} +   
\LLL_{\rm cap} + \LLL_{\rm ext}) / \rho .$
The first term is
\be
\LLL_{\rm adv}(\U) = -\grad \cdot ( \U  \U ) = -(\U \cdot \nabla) \U,
 \label{nsnonconsbis}
\nd
and the other terms \red {have been defined} above. 

\red {The conservation of momentum, on an elementary control volume moving with the 
interface $S$, leads to the following {\em jump conditions} across the interface}
\be
\centering
- \textlbrackdbl - p  +  2 \mu \,\N \cdot {\bf S} \cdot  \N \textrbrackdbl_S = \sigma \kappa  
\label{jump1}
\nd
\red {and} 
\be
\centering
- \textlbrackdbl 2 \mu \,\T^{(k)} \cdot {\bf S} \cdot  \N \textrbrackdbl_S = 
\T^{(k)} \cdot \nabla_{S} \sigma 
\label{jump2}
\nd
\red {where $\T^{(k)}$, $k=1,2$, are two independent tangent vectors
and the notation $\textlbrackdbl ... \textrbrackdbl_S$ denotes the jump
of a physical quantity across the interface $S$.
The Continuous Surface Force (CSF) \cite{brackbill92} formulation smoothens the 
surface tension force a few cells across the interface and does not require 
the jump conditions (\ref{jump1},\ref{jump2}), that are instead implemented in 
the Ghost Fluid Method (GFM) \cite{Fedkiw99}.} 


\subsection{Boundary conditions} 

A major difficulty with numerical simulations of fluid flow is the correct implementation
of the boundary conditions. In principle the conditions at boundaries are well defined.
For viscous, incompressible fluids we require that the fluid sticks to the wall so 
\red {that} the fluid velocity there is equal to the wall velocity
$$ \ubar = {\bf U}_{wall}.$$
In a numerical setup, we can also impose periodic boundary conditions, as well as
inflow or outflow conditions (see TSZ). 

\subsection{Free-surface flow} 
\label{subsec:fs}

Free-surface flow is a limiting case of flow with interfaces, in which the treatment 
of one of the phases is simplified. 
For instance, for some cases of air-water flow, we may consider the pressure
$p$ in the air to depend only on time and not on space (through, say,
some function $p_{\rm air}(t)$) and the viscous stresses 
in the air to be negligible.
The jump conditions (\ref{jump1},\ref{jump2}) become {\em boundary conditions} on 
the border of the liquid domain
\be
\left. \left( - p  +  2 \mu \, \N \cdot \red {{\bf S}} \cdot  \N \right) \right|_S
= \red {-} p_{\rm air} + \sigma \kappa  
\label{free_surf_norm_bc}
\nd
and
\be
2 \,\mu \,\left. \T^{(k)} \cdot \red {{\bf S}} \cdot  \N \right|_S  = 
\T^{(k)} \cdot \nabla_{S} \sigma \,.
\label{free_surf_kine_stress}
\nd 

\section{Numerical methods implemented in the code} 
\label{method}	

\subsection{Spatial discretization}

\begin{figure}
\begin{center}
\includegraphics[width=0.65\columnwidth]{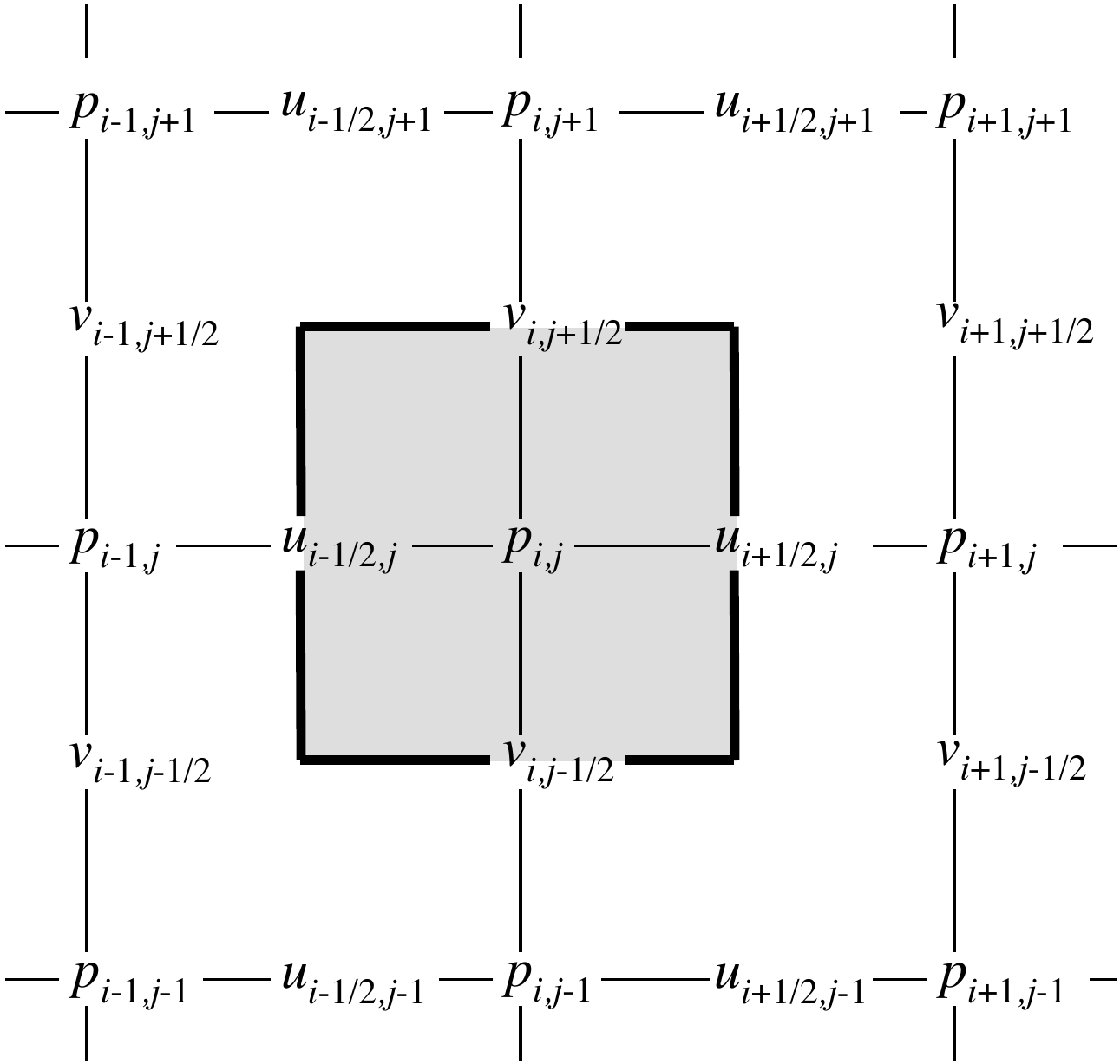}
\end{center}
\caption{Representation of the staggered spatial discretisation. The pressure $p$ is 
assumed to be known at the center of the control volume outlined by a thick solid line. 
The horizontal velocity component $u_1=u$ is stored in the middle of the left and right 
edges of this control volume and the vertical velocity component $u_2=v$ in the middle 
of the top and bottom edges.}
\label{stag-grid}
\end{figure}

We assume a regular cuboid grid, \red {that} can be defined as a cubic grid stretched
independently in the $x$, $y$ and $z$ directions, so that the centers of the cells 
$\Omega_{i,j,k}$ are given by the intersection set of planes $x=x_i,y=y_j,z=z_k$ and the 
cell boundaries are contained in the set of planes $x=x_{i+1/2},y=y_{j+1/2},z=z_{k+1/2}$. 
\red {However, in \pariss the Front Tracking method can indeed use stretched coordinates,
while the routines implementing the VOF method are still limited to cubic grids.}

\begin{figure}
\begin{center}
\begin{tabular}{cc}
  \includegraphics[width=0.65\columnwidth]{{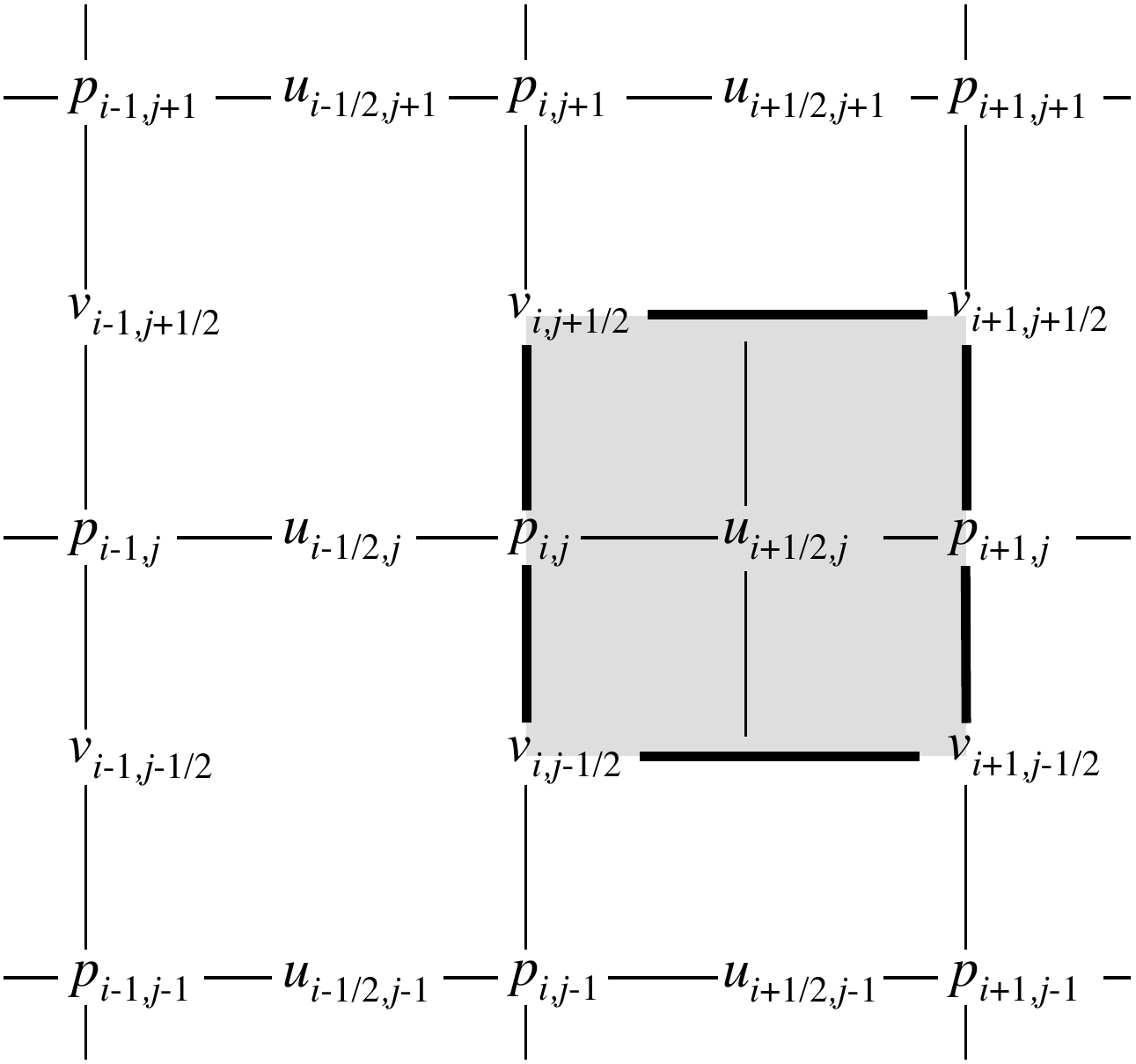}} & (a) \\ 
  \includegraphics[width=0.65\columnwidth]{{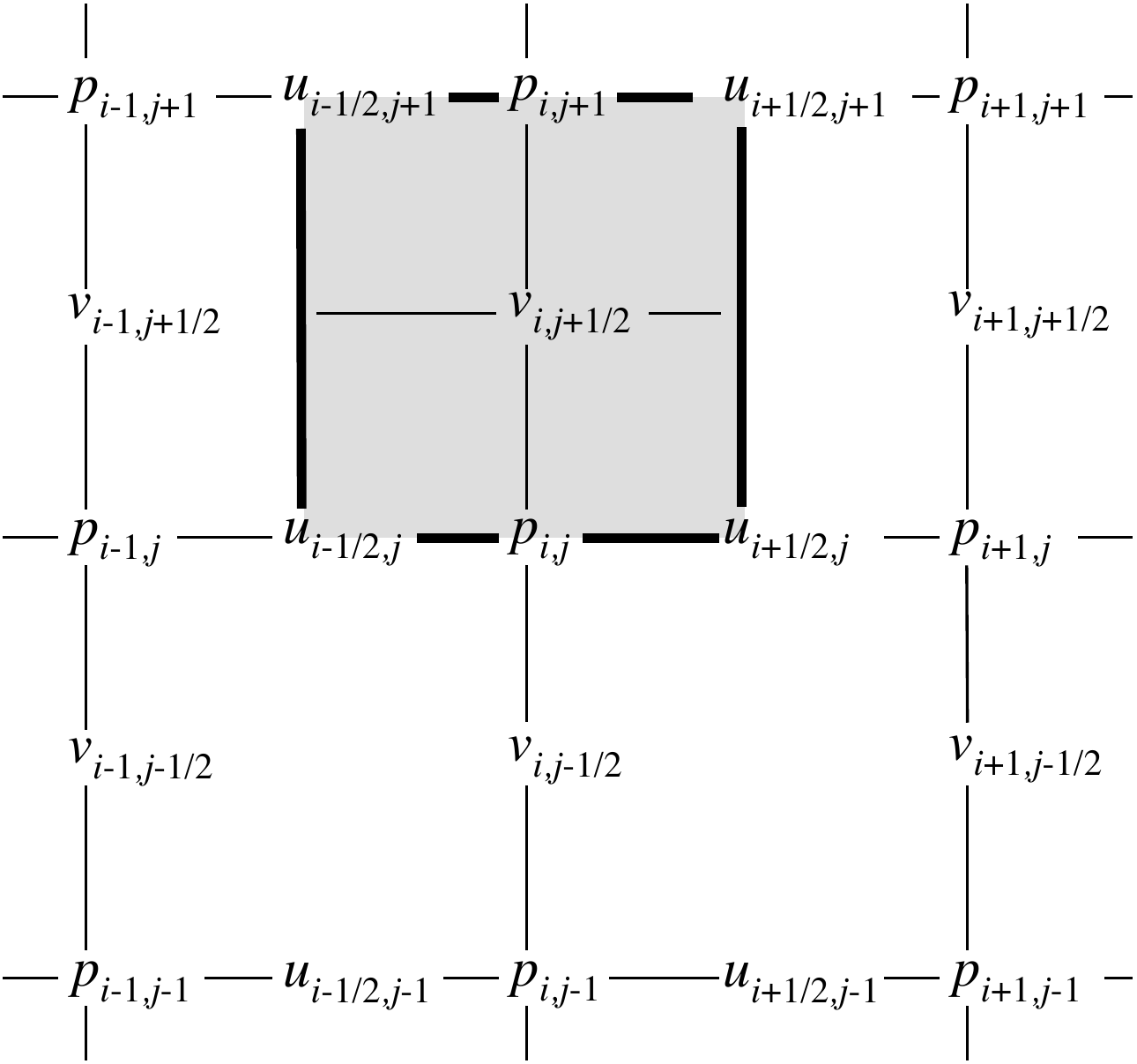}} & (b)
\end{tabular}
\end{center}
\caption{The control volumes for the $u_1=u$ and the $u_2=v$ velocity components are 
displaced half a grid cell to the right (horizontal velocities) and to the top 
(vertical velocities). Here the indexes show the location of the stored quantities. 
Thus, half indexes indicate those variables stored at the edges of the pressure 
control volumes.}
\label{Mac-u-v}
\end{figure}

\red {We use a finite volume discretization of the momentum equation and
consider staggered velocity and pressure grids. 
The staggered grid and control volumes for the pressure $p$
are represented in Fig. \ref{stag-grid}.
The corresponding control volumes of the velocity component $u_m$ in direction
$m$, $m=1,2,3$, are shifted with respect to the
control volume $\Omega_{i,j,k}$ surrounding the pressure $p$.
The control volumes for the velocity components $u_1$ and $u_2$ in two dimensions, 
or for the corresponding momentum components, are shown on Fig. \ref{Mac-u-v}.}
The use of staggered control volumes has the advantage of
suppressing neutral modes often observed in collocated methods but
leads to more complex discretizations (see \red {TSZ} for a
more detailed discussion.)  This type of staggered 
representation is easily generalized to three dimensions.

Using the staggered grid leads to a compact expression for the continuity equation
\eqref{divu}
\bea
\frac{u_{1;i+1/2,j,k}- u_{1;i-1/2,j,k}}{\Delta x} &+ &
\frac{u_{2;i,j+1/2,k}- u_{2;i,j-1/2,k}}{ \Delta y} \;\; + 
\nonumber \\ 
\frac{u_{3;i,j,k+1/2}- u_{3;i,j,k-1/2}}{ \Delta z}&=&0,
\label{cont2}
\nda
In what follows, we shall use the notation $f=m\pm$, with the integer index $m=1,2,3$,
to \red {denote} the face of any control volume located in the positive or negative 
Cartesian direction $m$, and $\N_f$ for the normal vector \red {to} face $f$ pointing
outwards of the control volume. On a cubic grid the spatial step is $\Delta x = 
\Delta y = \Delta z = h$ and Eq. \eqref{cont2} becomes
\be
\grad^h \cdot \U = \sum_{m=1}^3 (u_{m+} + u_{m-})/h =0 \,, 
\label{incompdisc3}
\nd
where $u_f = u_{m\pm}= \U \cdot \N_f $ is the velocity \red {component} normal 
to face $f$. 

It is worth noting that in the staggered grid setup, the control volume for $p$ is 
also the control volume for other scalar quantities, such as $\rho$, $\mu$ and $C$.

\subsection{Time Marching}

Time marching can be performed in a first-order or second-order manner using a
small, possibly variable time step $\tau$ so that $t_{n+1} = t_n + \tau$.
\red {We start with the description} of the first-order time stepping. 
The interface is first advanced in time as follows. 
\red {In the Front-Tracking method, the front points ${\bf x}_k$ are moved as}
\be
{\bf x}_k^{n+1} =  {\bf x}_k^{n} + \tau \, \U_i ({\bf x}_k^{n}, t_n)
\label{ptadv}
\nd
\red {where $\U_i(\X,t_n)$ is an interpolation of the velocity field at the location
$\X$ (see Section \ref{fronttracking} for the definition of the front points).
A marker function $I$ must then be constructed to compute the material
properties in the grid points, as described in Section \ref{markerf}.}

In the VOF method the \red {volume} fraction field is updated as
\be
C^{n+1} = \LLL_{\rm VOF}(C^{n},\U^{n}\tau/h) \,,
\label{cnp1}
\nd
where $\LLL_{VOF}$ represents the operator that updates the \red {C} data
given the velocity field $\U^{n}$ and is described in detail in Section \ref{vof}. 
\red {Once the volume fraction $C$ (or the marker function $I$) is updated, the 
material properties are computed from \eqref{muH}, while the interface geometry, 
including its unit normal $\N$ and curvature $\kappa$, is estimated with the
different methods described in the next sections. The velocity field is then 
updated with a predictor-corrector method. 
In the predictor step a provisional velocity field $\U^*$ is computed.}
Two different versions are used. \red {One is the ``non-consistent'', non 
conservative version of Eqs. (\ref{nsnoncons},\ref{nsnonconsbis})}
\bea
\U^* &=& \U^{n} +  \tau \LLL^h_{\rm adv}(\U^{n})  
\nonumber \\  
&+ &\frac{\tau}{\rho^{n+1}}  \big( \LLL^h_{\rm diff}(\mu^{n+1},\U^{n}) + 
\LLL^h_{\rm cap}(C^{n+1}) + \LLL^h_{\rm ext}(C^{n+1}) \big) \,, 
\label{nonconspredictedvel}
\nda
\red {the other is a ``mass-momentum consistent'', conservative version
of Eqs. (\ref{nscons},\ref{diff1})}
\bea
\rho^{n+1} \U^* & = & \rho^n \U^{n} + \tau \LLL^h_{\rm cons}(\rho^{n},\U^{n}) 
\nonumber \\
 &+ & \tau  \big( \LLL^h_{\rm diff}(\mu^{n+1},\U^{n}) 
+  \LLL^h_{\rm cap}(C^{n+1}) + \LLL^h_{\rm ext}(C^{n+1}) \big)\,.
\label{conspredictedvel}
\nda
Clearly the above operators depend on the discretization steps 
$\tau$ and $h$ as well as the fluid \red {properties}. 
The ``mass-momentum consistent'' \red {advection} operator  $\LLL^h_{\rm cons}$
is described in detail in \cite{fuster2018momentum}.
In the second step the pressure gradient {\em corrects} the velocity
\be
\U^{n+1} = \U^* - \frac{\tau }{\rho^{n+1}} \nabla^h p \,. 
\label{fotm}
\nd
\newcommand\pijk{\phi_{i,j,k}}
This step constitutes the so-called projection method. The pressure is determined by the
requirement that the velocity at the end of the time step must have zero divergence
\be
\grad^{h} \cdot \ubar^{n+1}=0 \,,
\label{cont-eq1}
\nd
which leads to an elliptic equation for the pressure
\be
\grad^h \cdot \frac{\tau }{\rho^{n+1}} \nabla^h p =  \grad^h \cdot \U^* \,. 
\label{poisson}
\nd
The whole set of operations above constitutes a first-order \red {approximation 
in time}, which can be written
\newcommand\interf{{\bf f}}
\be
(\interf^{n+1},\U^{n+1}) = \LL(\interf^{n},\U^{n}) 
\label{ts1}
\nd
where $\interf^n$ is the interface data (either the front points $\X$ or the VOF 
fraction $C$) and \red {the operator $\LL$ consists in the steps described above and 
is applied to the data at time $t_n$.}
A second-order time scheme can be obtained \red {by first computing a set of 
temporary variables $\interf^{**}$ and $\U^{**}$}
\be
(\interf^{**},\U^{**}) = \LL(\interf^{n},\U^{n})
\nd
\red {followed by the update of the variables at time $t_{n+1}$ 
by the trapezoidal rule
\be
(\interf^{n+1},\U^{n+1}) = \frac{1}{2} 
\big( \LL(\interf^{**},\U^{**}) + \LL(\interf^{n},\U^{n}) \big)
\nd
where the operator  $\LL$ on the right hand side effectively is applied
to data at the intermediate time $t_{n+1/2}$. The \pariss code implements both 
the first-order and the second-order time schemes}, controlled by the parameter 
{\sf ITIME\_SCHEME}.

\newcommand{\dpt}[1]{\frac{\partial #1}{\partial t}}

\subsubsection{Non-conservative momentum advection}

The non-conservative momentum advection in Eqs. (\ref{nsnoncons},
\ref{nsnonconsbis}) amounts to integrate over one time step the PDE
\be
\dert u_m = \LLL_{{\rm adv}, m}(\U) 
\nd
where
\be
\LLL_{{\rm adv}, m}=  \red {-} u_j \partial_j u_m 
\nd
for each value of the index $m$. Because of incompressibility (\ref{divu}) 
it is equivalent to solve for a scalar field $\phi = u_m$ in the manner
\be
\dert \phi + \nabla \cdot ( \phi \U)  = 0 
\label{phiconv}
\nd
Integrating the advection equation \red {\eqref{phiconv} in space,} over a control
volume centered on a node of the scalar $\phi$, and in time one obtains
\be
{\pijk^{n+1} - \pijk^{n}} = - \sum_{\rm{faces}\, f} F^{(\phi)}_f. 
\label{sumfp}
\nd
We use $ F^{(\phi)}_f = \phi_f \U_f \cdot \N_f \tau/h$ as an approximation of 
the flux on face $f$.
Let $u_f = \U_f\cdot \N_f$. At first order \red {the advecting velocity component}
$u_f$ is obtained by simple averaging \red {or centered interpolation}. 
\red{For example, we can consider the first component of velocity $\phi=u_1=u$,
whose control volume is centered at $i+1/2,j,k$ (see Fig. \ref{Mac-u-v}).
For the flux along the horizontal $x$ direction,  
the right face is located at index $i+1,j,k$. For the normal component of the 
velocity on this face, the interpolated value is 
$u_f = u_{1;i+1,j,k} = \frac 12 (u_{1;i+1/2,j,k} + u_{1;i+3/2,j,k})$. 
On the other hand for the flux along the $y$ direction, the top face is located 
at index $i+1/2,j+1/2,k$ and the advecting velocity component is now 
$u_f = u_{2;i+1/2,j+1/2,k} = \frac 12 (u_{2;i,j+1/2,k} + u_{2;i+1,j+1/2,k})$.}

Contrary to the estimates of $u_f$ above, the estimation of \red{the advected 
quantity} $\phi_f$ may involve more complex and higher-order schemes. Indeed
one-dimensional interpolation schemes incorporating flux limiters are typically used. 
Most of these schemes are described in TSZ together with their usage in computing 
the face fluxes in the bulk. We describe their general properties here shortly. 
Since the schemes are one-dimensional we can consider a variable $\phi$ defined 
on a regular one-dimensional grid. Specializing further the example, we  
\red {assume} that the variable $\phi$, akin to the \red {velocity} component $u_1$, 
takes values  $\phi_{i+1/2}$ at half-integer grid point indexes. 
We need to estimate the flux at integer index points $x_i$, \red{thus we need} 
to predict $\phi_i=\phi_f$. \red {To this aim,} an interpolation function is defined 
that \red {computes} this value as a function \red {of the value of $\phi$ at} the 
four nearest points, and in an upwind manner based on the sign of \red {$u_i=u_f$}, 
that is given by the \red {previously-defined} centered interpolation.  The face value 
$\phi_i$ is then approximately given by 
\be 
\phi_i = f(\phi_{i-3/2}, \phi_{i-1/2}, \phi_{i+1/2},\phi_{i+3/2},u_i) 
\nd 
where the function $f$ is both of sufficiently high order and \red {such as to} limit 
the flux. To express the function $f$, \pariss offers a choice of the ENO, QUICK, 
Superbee, WENO, first-order upwind, Verstappen, or BCG schemes.

When momentum is advected with the \red {non conservative formulation
\eqref{nonconspredictedvel}}, $\phi$ in Eq. (\ref{sumfp}) 
is taken equal to one of the velocity components $u_m$. This method is 
available whether 
one uses the VOF method or the Front-Tracking method. \red {With the VOF method
we have extensively used a combination of QUICK, away from the interface, and
a first-order upwind near the interface, and a combination of Superbee slope limiter,
away from the interface, and its modified version near the interface. 
These combinations have been found numerically more robust than others, and
several simulation results are presented in \cite{fuster2018momentum}.

With the Front Tracking  method we use typically ENO or QUICK, as they offer a good combination of accuracy and stability.
}

\subsubsection{\red{Mass-momentum} consistent momentum advection}

When using VOF, another momentum advection method is available that is consistent with VOF
advection and which implements a conservative scheme of the form (\ref{conspredictedvel}).
This means that the same advection method is used near the interface for \red {both} the 
VOF \red {volume fraction} $C$ and the velocity $\U$. In other words, when there is a 
density jump on the interface, the discontinuity of the \red {momentum density} $\rho \U$ 
is advected exactly at the same velocity as the discontinuity of the 
\red {mass density $\rho$}. This can be expressed by saying that the momentum advection 
and the VOF advection are {\em consistent}. An explicitly formulated criterion for 
consistency is the following: if the velocity $\U$ is
uniform, then $\rho \U$ remains exactly proportional to $\rho$. This should
happen even \red {when} $\rho$ is obtained from
the VOF-advection of $C$ using Eqs. (\ref{muH}) and (\ref{cnp1}) 
 and $\rho \U$ is obtained from the operator $\LLL^h_{\rm cons}$. 
Such a ``VOF-consistent'' method is used since it has been empirically found by several 
authors that the \red {non-consistent} advection of the previous section was often 
unstable at large density ratios, while consistent methods are more
stable \cite{bussmann2002modeling,desjardins10,raessi12,le13,ghods2013consistent,
Vaudor:2017ip,patel2017novel,nangia2019robust,ivey17,owkes17}.

The \pariss code implements a modification of the classical momentum-preserving scheme 
proposed by \cite{rudman98} for the case of a staggered grid and Volume of Fluid (VOF) 
method. The scheme needs to be modified from the one in the previous section only near 
the interface. In particular, away from the interface, the density is constant and the 
scheme in (\ref{sumfp}) is already conservative. 
\red{Near the interface, with 
$\phi =  \rho u_m$, the momentum advection can be written, see 
\cite{fuster2018momentum} for details, as
\be
\big( \rho u_m \big)^{n+1}_{i,j,k} - \big( \rho u_m \big)^{n}_{i,j,k} = 
- \sum_{\rm{faces}\, f} u_{m,f}\, F^{(\rho)}_f \,+\, \sum_{l=1}^3 
\widetilde u_{m}\, C^{(\rho)}_l \,,
\label{sumfpu}
\nd
where $u_{m,f}$ is the one-dimensional interpolation of the velocity component
$u_m$ on face $f$ and $F^{(\rho)}_f$ is the mass flux through the same face.
The last sum represents the compressional term and it is related to the fact that 
each term in Eq. \eqref{incompdisc3} can be different from zero even if the flow 
has zero divergence. In particular this sum is equal to zero for the mass-conserving 
VOF method of \cite{Weymouth:2010hy}.

The one-dimensional interpolation schemes are different if they are applied
away from or near the interface. The combinations that are used with the 
mass-momentum consistent advection have been already discussed in the previous
section.}

\subsubsection{Implicitation of the viscous terms}

The operator $\LLL^h_{\rm diff}(\mu^{n+1},\U^{n})$ may be treated in part implicitly. 
\red {From (\ref{diff1},\ref{diff2})} we have \red{
\be
\LLL_{\rm diff, j}(\mu,\U) = (\grad \mu) \cdot \big( \nabla\U + (\nabla\U)^T \big)
+ \mu \nabla^2 \U  
\nd
}
The first term on the RHS is left explicit, but the second term can be made implicit 
by solving the linear problem\red{
\be
\U^{*} = \U^{n} +  \tau \,\mu^{n+1} \nabla^{2} \U^*\, .   
\label{muelliptic}
\nd
}
Then the discrete operator is defined as \red{
\be
\LLL_{\rm diff, j}^h(\mu^{n+1},\U^n) =  
(\grad \mu^{n+1}) \cdot \big( \nabla\U^n + (\nabla\U^n)^T \big)
 + \big( \U^{*} - \U^{n} \big) \big/ \tau
\nd 
}
where $\U^*$ is the solution of the linear problem (\ref{muelliptic}). 
The implicitation of the viscous terms is optional and controlled by a code parameter.

\subsection{Interface advection: Front-Tracking method} 
\label{fronttracking}

\begin{figure}
\begin{center}
    \includegraphics[width=0.95\columnwidth]{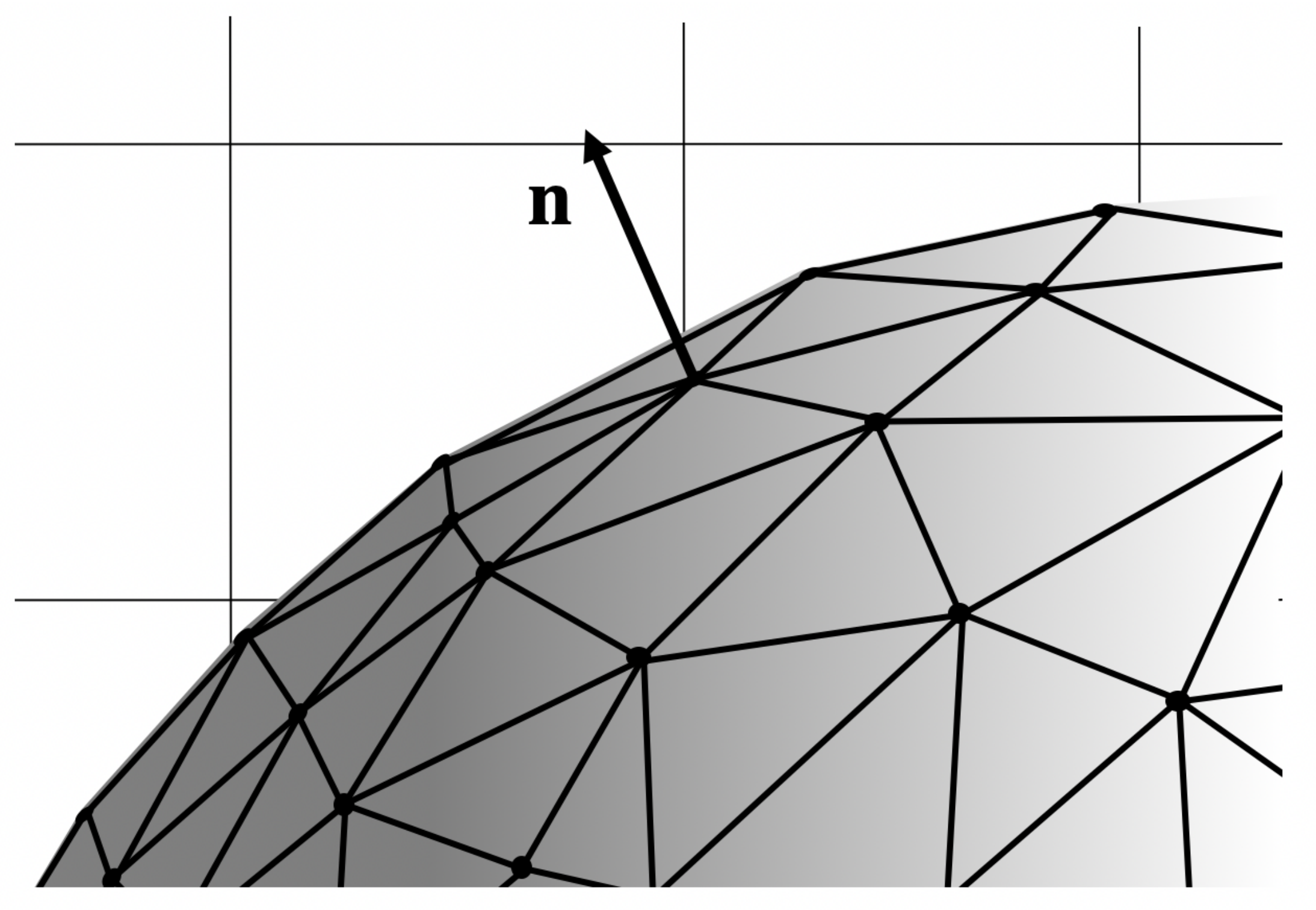}
\end{center}
\caption{An interface separating two immiscible fluids represented by marker points 
connected by triangular elements.}
\label{FrontFig1}
\end{figure}

\red {The interface advection in the ``one-fluid'' formulation is required to
update the material properties, see Eq. \eqref{muH}, and the interface geometry
in order to compute the surface tension forces. In the \pariss code the interface
 can be moved} either by Front Tracking (FT) or by 
the VOF method. We describe the former in this section. \red {The FT method},
in the context of simulations of two or more immiscible fluids, refers to tracking the 
interface separating the different fluids using moving connected marker points that 
represent the interface. In our implementation the marker points are connected by triangular 
elements, where the points are ordered in the same way for all elements, allowing us 
to define an ``inside'' and an ``outside'' for each element. The coordinates of the points 
are stored in arrays, in arbitrary order, with separate integer arrays providing
pointers to the previous and next points. Thus, the points form a linked list where the 
location in the array provides each point with a unique ID. 
The elements are stored in the same way, with arrays containing pointers to the corner or 
node  points. The coordinates of the marker points are the main quantities stored for 
the points, but the points also have arrays for various temporary quantities, such as
velocities and the surface force. The marker points and the elements connecting the
points together form the ``front.'' In addition to pointers to their corner points, 
the elements also have pointers to the elements that share edges with them. 
These are mostly used for modifications, or reconstruction, of the front. Notice that 
generally only one front is needed, irrespectively of the number of distinct interfaces, 
and that distinct interfaces, \red {as present in a simulation containing several bubbles 
and drops, can have different material properties, for example surface tension.} 
Figure \ref{FrontFig1} shows the layout of a triangulated front separating two different 
fluids.

\begin{figure}
\begin{center}
    \includegraphics[width=0.95\columnwidth]{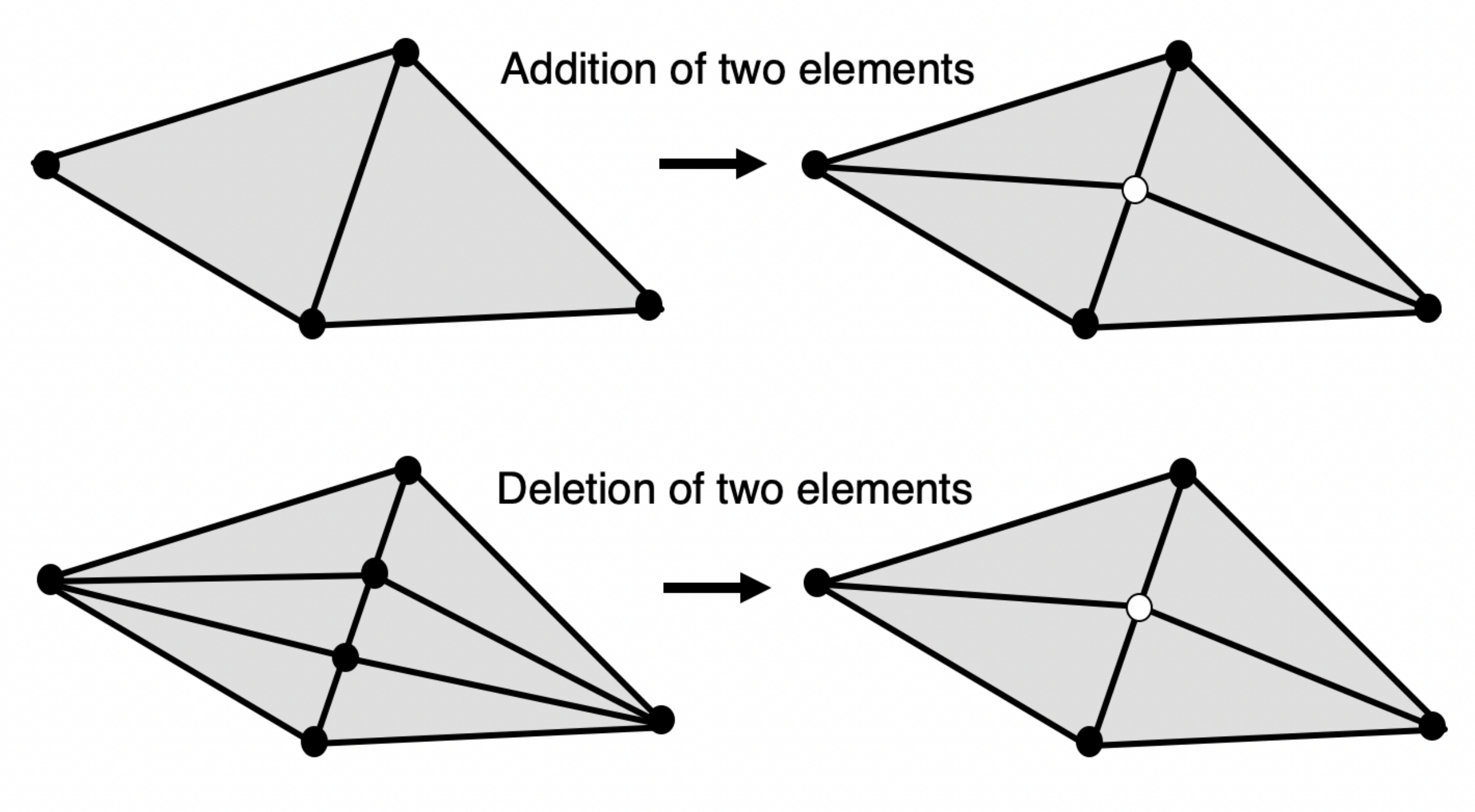}
\end{center}
\caption{Restructuring of a triangulated grid by adding and deleting points and elements.}
\label{FrontFig2}
\end{figure}

As the front is deformed, stretched and compressed by the flow, the size of the elements 
\red {changes} as points move away from, or closer to, each other. We keep the length of 
the edges of each \red {element} within about a quarter to a half of the
\red{spacings of the fixed fluid grid}, and to maintain that resolution, points and 
elements are dynamically added and deleted. While many strategies are possible, we add 
points by splitting the longest edge of an element by adding one point and two 
elements, and delete points by collapsing the shortest edge of an element, removing one 
point and two elements. The grid quality can sometimes also be improved by changing the 
connectivity of the elements but we generally find that doing so is not necessary. 
The addition and deletion of front points and elements is shown in Fig. \ref{FrontFig2}.

\red {Topology changes in front tracking can be done in several slightly different ways, but in 
all cases additional code is required to: (a) detect points where coalescence/breakup 
should take place, and (b) restructure the front to account for the topology change. 
For a short description of a topology change algorithm and a few examples see \cite{razi18}
and for applications of a slightly modified version of the algorithm to more complex flows 
see \cite{lutry18} and \cite{lutry19}. A detailed description of a topology change algorithm 
is beyond the scope of the current manuscript.} 

\subsubsection{Connecting the front and the fluid grid}

Since the Navier-Stokes equations are solved on a fixed grid, we have two grids: the 
front/interface grid and the fixed grid.
\red {The} motion of the interface depends on the flow and the flow depends
on where the interface is, \red {therefore} information must be passed back and 
forth between the front and the fixed fluid grid. 
To do so we need to identify what front point is close to which fixed grid point
and vice versa.  For a regular structured grid, where the grid lines are straight and 
evenly spaced, it is straightforward to locate a point on the fixed grid that is closest 
to a given front point (using an INT or a MOD function), but finding the 
front point closest to a given grid point generally requires us to examine the distance 
to all the front points.
Thus, it is more efficient to do all communications between the front and the fixed 
volume grid by looping over the front points. For periodic domains we allow the front 
to move out of the domain resolved by the fixed fluid grid, and use a MOD function to 
find the fixed grid point that would be closest to a given front point if we moved the 
front back into the original domain. 

To transfer information between the fixed fluid grid and the moving front, we need 
\red {on one side to interpolate data} from the grid \red {to the front 
and on the other side to} spread, or ``smooth,''  \red {data} from the 
front to the fixed grid. \red {The first type of data transfer is required to
move the front, where the velocity at the front points is 
interpolated from the velocity on the} fixed grid. On a staggered grid each 
velocity component is interpolated separately. In general, we have
\begin{equation}
\phi^l_f=\sum_{ijk} w^l_{i,j,k} \,\red {\phi_{i,j,k}}
\label{Interpolation1}
\end{equation}
Here, $\phi^l_f$ is a quantity, such as the velocity, on the front at point $l$, 
\red {$\phi_{i,j,k}$} is the same quantity on the fluid grid, $w^l_{i,j,k}$ is the weight 
of each grid point with respect to front point $l$ and  the sum is over grid points 
``close'' to the front points. Generally the same time integration method is used for the 
advection of the points as is used for updating the fluid velocities. 

\red {The second type of data transfer is usually referred to as smoothing,
since it replaces a quantity defined at a front point on the sharp interface, 
such as surface tension}, with a distribution on the fixed grid, where each 
fixed grid point receives a value according to how close it is to the front point. 
Unlike the interpolation of a quantity\red {,} such as the velocity from the fixed grid 
to a front point, smoothing usually involves quantities like a force, that are given in 
terms of force per unit interface area on the front but must be converted to force per 
unit volume on the fixed grid, so that the total force is conserved. Thus, the quantity 
smoothed must be scaled by the ratio of the area associated with each front point 
divided by the volume of a fixed grid cell 
\begin{equation}
\red {\phi_{i,j,k}} = \sum_{l} \phi_f^{l} \,w_{i,j,k}^l \,{\Delta \red {A}_{l}\over 
{\Delta x \Delta y \Delta z}},
\label{Interpolation2}
\end{equation}
where $\Delta \red {A}$ is a surface area of a front element and $\Delta x$, $\Delta y$ 
and $\Delta z$ are the grid spacings.

Several interpolation/smoothing functions can be used, but in the PARIS code we use a 
smoother interpolation function originally introduced in \cite{Mcqueen:89}
which involves four grid points in each coordinate  direction, or 64 points total. 
The weights are given by 
\begin{equation}
w^l_{i,j,k} = d(r_x) \, d(r_y) \, d(r_z), 
\label{Weights}
\end{equation}
where $r_x$ is the scaled distance (by the grid spacing) between $x^l_f$ and the grid 
line located at $x_i$\red {;} $r_y$ and $r_z$ are defined in the same way.  
In our case, \red {the weighting function $d(r)$} is given by
\begin{equation}
d(r)=\left\{\begin{array}{ll}
   (1/4)(1+\cos(\pi r/2)), & \left| r \right| <2 ,\\
    0, &  \left| r \right| \geq 2.
   \end{array} \right. 
\label{Peskin1}
\end{equation}
Using fewer points gives a sharper transition  zone but sometimes leads to wiggles, 
particularly for stiff problems. The interpolation function is bounded with weights 
that sum to one in addition to having various desirable symmetry properties. 
For a discussion see the reference above. 



\subsubsection{Constructing the marker function \red{I}}
\label{markerf}

Once the front has been moved, a marker function must be constructed on the fixed grid 
to assign the different material properties to each grid point. This can be done in many 
different ways, but one of the consideration is that fronts that are so close to each 
other that the flow between them is not resolved, must be handled in a plausible way. 
Usually this means that the marker function must retain its correct value on both sides 
of the double front. In the PARIS code we do this by working with the gradient of the 
marker function \red{I}, which in the limit of a sharp interface should be a delta function 
defined only on the interface. The delta function is then treated in the same way 
as the surface tension and smoothed onto the fixed grid. 
\red {The grid value of the components of the gradient is given by Eq. 
\eqref{Interpolation2} of the previous section
\begin{equation}
\big( \grad \markerfunction \big)_{i,j,k} = \sum_{l} \big( \Delta \markerfunction \,\N 
\big)_l \,w_{i,j,k}^l \, {\Delta \red {A}_{l}\over {\Delta x \Delta y \Delta z}} \,,
\label{gradI}
\end{equation}
where $\Delta \markerfunction$ is the jump in the value of the marker function across 
the interface (usually a given constant value) and $\N$ the unit normal to the element. 
Once the gradient field has been} smoothed
onto the fixed grid, we can integrate it to recover the marker function \red {I. 
For example, on a staggered grid}
\begin{equation}
\markerfunction_{i,j,k}=\markerfunction_{i-1,j,k}+{\Bigl( {\partial \markerfunction 
\over \partial x} \Bigr) }_{i-1/2,j,k} \Delta x \,.
\label{FrontMarker1}
\end{equation}
To maintain symmetry, the marker at a given point is constructed as the average of 
the integration from all the neighboring points, leading to a linear system that is solved 
iteratively. Using standard second-order centered finite-difference approximations, 
the linear system is an approximation to
\begin{equation}
\nabla^2 \markerfunction = \nabla \cdot (\nabla \markerfunction)_f,
\label{FrontMarker2}
\end{equation}
where the subscript on the gradient of $\markerfunction$ on the right hand side means that 
it comes from the front. Since the right hand side is known, as it is deduced from the 
position of the front through (\ref{FrontMarker1}), this equation 
amounts to the Poisson equation
\begin{equation}
\nabla^2 \markerfunction = \nabla \cdot \mathbf{G}_f
\label{FrontMarker3}
\end{equation} 
which must be solved to find $\markerfunction$ \red {for a given $\mathbf{G}_f = \big( \grad \markerfunction \big)_f$}.
In the current version this equation is solved for the entire grid \red {to keep 
the implementation simple}, but this can easily be 
changed to involve only grid points next to the interface, where the value of the marker 
function changes as the front moves. Integrating the gradient of the marker function on 
the fixed grid is, of course, only one way to construct it. We have, however, found that 
doing so generally leads to a smooth but compact transition from one fluid to the other. 
In addition, since the gradients of two interfaces bounding a very thin film cancel each 
other when transferred to the fixed grid, the marker there will ``disappear''. This seems 
like the proper way to treat films too thin to be resolved on the fixed  grid.
\begin{figure}
\begin{center}
    \includegraphics[width=0.95\columnwidth]{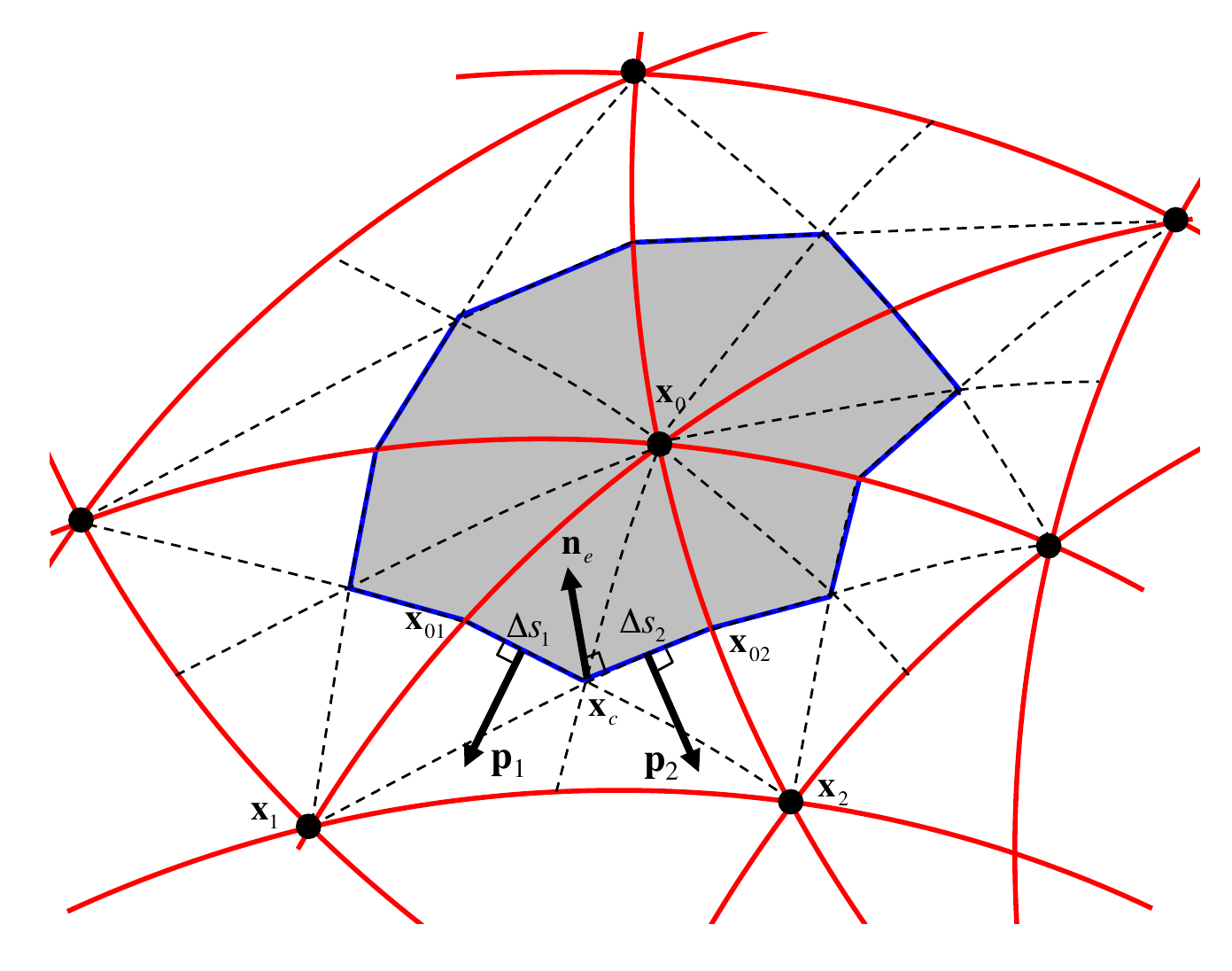}
\end{center}
\caption{Computation of the surface force on a triangulated grid by integrating over 
the edges of an element.}
\label{FrontFig3}
\end{figure}

\subsection{\red {Surface tension: Front-Tracking method}} 
\label{fronttension}

In simulations of flows with sharp interfaces the front serves two main functions. The 
first is the advection of the marker function \red {I}, as described above, and the second is 
the computation of the surface tension described below.  As with most of the other operations 
for the front, finding the surface force can be done in several different ways. In the PARIS 
code we compute the force on the front and transfer it to the fixed fluid grid, where it is 
added to the discrete Navier-Stokes equations.

To ensure that the force is conserved as we transfer it from the  front and onto the fixed 
volume grid, we work with the total force on a small area.  The total force on a small 
region surrounding a front point is computed by dividing each front element into three 
equal parts, each connected to  one nodal point, and computing the pull on their side as 
described below. The force from each part is then added to the appropriate nodal point. 
When the surface force on all the elements has been computed, it is transferred to the 
fixed grid and converted into a force per unit volume by (\ref{Interpolation2}). 
Working with the total force on a surface element, rather than the force per area, makes it 
easier to ensure the total force is conserved when it is transferred between the front and 
the fixed volume grid.

To find the surface force we use the fact that the total force on a surface element can be 
found by integrating the ``pull'' on its edges \red {
\begin{equation}
{\bf f}_{\sigma}  = \int_{\Delta A} \sigma \, \kappa \, \N_e \,d A = 
\oint_{C} \sigma \,{\bf p} \,d l
\label{FrontSurfTens1}
\end{equation}
Here the Stokes theorem has been used to convert the area integral over $\Delta A$
into a line integral along its boundary $C$. The unit vector 
${\bf p} = {\bf t} \times {\bf n}_e$ is tangent to the interface and perpendicular to the 
boundary of the interface element (see Fig. \ref{FrontFig3}).}
By keeping the surface tension coefficient $\sigma$ under the integral sign we allow for 
variable surface tension. The benefit of using this expression is that we only need to 
approximate tangents on the surface, not the curvature, and that the pull on the side of 
one surface element is equal and opposite to the pull on the adjacent element. 
Thus, the surface force is conserved in the sense that an integral over a surface patch 
consisting of several surface elements is \red {guaranteed} to give the same \red {result} as 
an integral over the boundaries of the whole patch. For constant surface tension coefficient, 
the integral over a closed surface is, in particular, guaranteed to be zero. 

Figure \ref{FrontFig3} shows schematically how the integration is done. We assume that the 
elements are flat and that surface tension, given at the front points, can be non constant.  
It therefore has to be interpolated when approximating the integral. 
The force at point ${\bf x}_0$ is computed as the ``pull'' on the edges of the gray patch, 
surrounding it. The contribution from the element connecting points ${\bf x}_0$, ${\bf x}_1$ 
and ${\bf x}_2$  is found by first splitting it in three \red {parts} by connecting the 
centroid ${\bf x}_c = (1/3) ({\bf x}_0+{\bf x}_1+{\bf x}_2)$ to the \red {midpoints} of the edges 
${\bf x}_{01} = (1/2) ({\bf x}_0+{\bf x}_1)$ and 
${\bf x}_{02} = (1/2) ({\bf x}_0+{\bf x}_2)$ and then finding the force on those edges by 
approximating the integral using a \red {midpoint} rule
\begin{equation}
\Delta {\bf f}_{\sigma} \approx \sigma_1 \Delta s_1 {\bf p}_1 + \sigma_2 \Delta s_2 
{\bf p}_2 \,.
\label{FrontSurfTens2}
\end{equation}
Here, the pull on each element is the cross product of the outward unit normal, ${\bf n}_e$, 
and the tangent vector to the edge, i.e. $\Delta s_1 {\bf p}_1 = {\bf n}_e  \times 
({\bf x}_{01} - {\bf x}_c) $ and $\Delta s_2 {\bf p}_2 = {\bf n}_e \times ({\bf x}_c - 
{\bf x}_{02})$,  where $\Delta s_1$ and $\Delta s_2$ are the lengths of the edges, and 
$\sigma_1$ and $\sigma_2$ are the surface tensions at the midpoint of the edges. The normal 
$ \red {{\bf n}_e}$ is found by the normalized cross product of two of the tangent vectors to 
the edges of the element. After straightforward algebra, we have
\begin{equation}
\Delta {\bf f}_{\sigma} \approx \frac{{\bf n}_e}{3} \times \Big ( 
\sigma_{1} \big(  {\bf x}_2 -{1 \over 2}( {\bf x}_1 + {\bf x}_0) \big) -
 \sigma_{2}   \big( {\bf x}_1 -{1 \over 2}( {\bf x}_2 + {\bf x}_0) \big) \Big)
\label{FrontSurfTens3}
\end{equation}
where the interpolated surface tension at the midpoint of the edges is
\begin{equation}
\sigma_{1}={1 \over 2} \Big( \frac12 \big( \sigma ({\bf x}_0)+\sigma ({\bf x}_1) \big) +
\frac13 \big( \sigma ({\bf x}_0)+\sigma ({\bf x}_1)+\sigma ({\bf x}_2) \big) \Big)
\nonumber
\end{equation}
\begin{equation}
= {1 \over 12} \big( 5 \sigma ({\bf x}_0)+ 5 \sigma ({\bf x}_1) +2 \sigma ({\bf x}_2) 
\big) 
\end{equation}
 and $\sigma_{2}$ is given by a similar expression.
The forces from the other elements connected to point ${\bf x}_0$ are found in the 
same way and added to give the total force on the \red {nodal} point, that is then 
``smoothed'' onto the fixed grid. 


\subsection{Interface advection: VOF method}
\label{vof}
When the interface location is \red {captured} by the VOF method, a variable $\cijk$ is 
initialized. It is equal to the fraction \red {of the cell $\Omega_{i,j,k}$ that is filled 
with the reference fluid $1$. The implementation of the VOF method is limited for the time 
being to cubic cells of edge length $h$. Moreover, we use a rescaling of space and time 
variables so that the cell size and the time step are both $1$}. All velocities are then 
rescaled to $u^\prime = u \tau / h$. Because of this space rescaling and in these new units, 
$\cijk$ is also the measure of the volume of reference fluid in cell $i,j,k$.

\subsubsection{Normal vector determination} 

The VOF method proceeds by a sequence of \red {interface} reconstructions and advections 
of $C$. In the reconstruction step, one attempts to \red {recover} the interface geometry 
from the VOF data $\cijk$. In the \pariss code we use already-published methods (see for 
example TSZ) that have been experienced to work satisfactorily. One first determines the 
interface normal vector $\N$, then solves the problem of finding a plane perpendicular 
to $\N$ \red {which cuts the cube with} exactly the volume $\cijk$. In \pariss two methods 
exist for normal vector determination. The most frequently used is the 
Mixed-Youngs-Centered Scheme (MYCS) \red {\cite{aulisa07}}, described also in TSZ. 
However, when a quick determination of the normal is needed \red {but not very accurate}, 
the finite difference method $\N = \nabla^h C$ (the Youngs scheme \red {\cite{youngs84}}) 
can also be used. 

\subsubsection{Plane constant determination}

Once the the interface normal vector $\N$ is determined, a new, colinear normal vector 
noted $\M$ and having unit ``box'' norm is deduced from $\N$, that is 
$||\M||_1 = |m_x| + |m_y| + |m_z| = 1$. Considering the volume $V=\cijk$ in cell $i,j,k$ 
the plane constant $\alpha$ is defined so that the plane 
\be 
\M \cdot \X = \alpha 
\nd 
cuts exactly a volume $V$ of the \red {cube}. The origin of the coordinate system
is taken at the corner of cubic cell $i,j,k$ with the smallest
coordinate values. The reader is reminded that we \red {are using} rescaled units
of space, so that $0 \le V \le 1$ and $0 \le \alpha \le 1$. 
Then $\alpha$ is determined by the resolution of a cubic equation \cite{Scardovelli00}. 
This resolution, and similar \red {ones often used in the VOF method and derived
in \cite{Scardovelli00},} are implemented 
in a kind of small library contained in the single file {\sf vof\_functions.f90}. 

\subsubsection{Volume initialization}
\label{vofi}

Before any VOF interface tracking is performed, the field of $\cijk$
values must be initialized. The \pariss code avoids inaccurate
initializations that for example initialize a sphere as a set of
$\cijk$ values which are \red {either} $0$ or $1$, a so called ``staircase'' or
``lego'' initialization. There are two ways in which initialization
can be improved over the lego one.  In the ``subgrid'' initialisation,
the mesh cells are \red {subdivided} into $n_I^3$ subcells (where $n_I$ is a tunable
parameter, called {\sf REFINEMENT} in the code). Then a ``lego''
initialization is performed trivially in the subcells. For example, if
the initial interface is defined implicitly by the equation $\phi(\X) = 0$ 
where $\phi$ is a smooth implicit function (akin to a level-set
function) then the trivial ``lego'' initialization \red {in each subcell
is $c_{I} = \Heaviside \big( \phi(\X_c) \big)$, where $\Heaviside$ is the Heaviside 
function and $\X_c$ the subcell center. The cell value $\cijk$ is then given 
by the sum of the $c_{I}$ values divided by the subcells number $n_I^3$.}
In tests it was found that $n_I=8$
was sufficient. However $n_I \ge 8$ leads to a very large number of
evaluations of the function $\phi$ and a slow initialization. In order
to avoid this, the \pariss code may be linked to the {\sc Vofi} library
described in \cite{bna2015numerical} and \cite{bna2016vofi}. Then the
initialization is performed using \red {highly-accurate} numerical 
integration \red {to compute the fluid volumes in each grid cell},
implicitly defined by the 
equation $\phi(\X) < 0$. \red {The code is linked to the {\sc Vofi} library} 
with the shell variable {\sf HAVE\_VOFI} set before compilation. 

\subsubsection{General split-direction advection}

\red {The reconstruction at time $t_{n}$ provides the approximate position of 
the interface, which is used to compute the reference phase fluxes across the cell 
boundary in order to update the volume fractions $\cijk$ at time $t_{n+1}$.} 
The \pariss code contains \red {two methods for split-direction advection}, which 
can be selected by the user: Lagrangian Explicit (LE) advection, with the keyword 
{\sf VOF\_ADVECT = LE}, and Weymouth and Yue (WY) advection, with the keyword 
{\sf VOF\_ADVECT = WY}. The LE advection is also called  ``Calcul d'Interface 
Affine par Morceaux'' (CIAM) which is \red {French} for ``Piecewise Linear 
Interface Calculation'' (PLIC)\red {, however} PLIC refers to generic VOF methods with a
piecewise linear reconstruction step, while CIAM refers to a specific type
of advection method first described in the archival literature in
\cite{li95} and classified as the ``LE'' method in \cite{Scardovelli02}.
The main advantage of both LE and WY is that they avoid overshoots ($\cijk > 1$)
and undershoots ($\cijk< 0$). Moreover WY conserves mass to machine accuracy. 
These methods are described in detail in TSZ for LE/CIAM and in 
\cite{Weymouth:2010hy} for WY. The reader may also refer to 
\cite{fuster2018momentum} for a condensed description of both methods. 

An important operation \red {in some simulations is the ``clipping'' procedure, 
and a small parameter $\epsilon_c$ is defined to this purpose}. After advection,
all cells that have \red {$\cijk < \epsilon_c$} are set to 0 and all cells that 
have \red {$\cijk > 1 - \epsilon_c$} are set to 1. This removes some, but not 
all, of the wisps, floatsam and jetsam \red {that are generated}. In the current 
version of the code the default value is $\epsilon_c = 10^{-8}$. \red {This is a 
rather high value used mainly with WY in simulations at very high density
contrast. The code has been observed to run well also with the smaller 
threshold value $\epsilon_c = 10^{-12}$ when the ratio of physical parameters 
is less extreme, or with no threshold at all with the LE advection. Other VOF schemes,
and in particular unsplit schemes, are not reported to require clipping.}

\subsection{Surface tension: VOF method}  
\label{surfacetension}
\subsubsection{CSF method}
\label{sec-csf}

For simplicity, we consider only the case where $\sigma$ is constant. 
In the Continuous Surface Force (CSF) method \cite{brackbill92} \red {the 
force $\sigma \kappa \N \delta_S$ in the capillary term (\ref{lllcap}) 
is written as}
\be
- \sigma \kappa \nabla \Heaviside = - \sigma \kappa^h \nabla^h C. 
\label{csf2}
\nd
where we have used the properties of the Heaviside function $\Heaviside$.
One of the advantages of this formulation is that it is a ``well-balanced'' 
method (see TSZ, Chapter 7 or reference \cite{popinet2018numerical}). 

An approximation for $\kappa$ needs to be found to use the CSF method.
A good estimate is obtained using so-called height functions. 

\subsubsection{Height Functions}
\label{height_function}

\begin{figure}
\begin{center}
\includegraphics[width=0.95\columnwidth,angle=0]{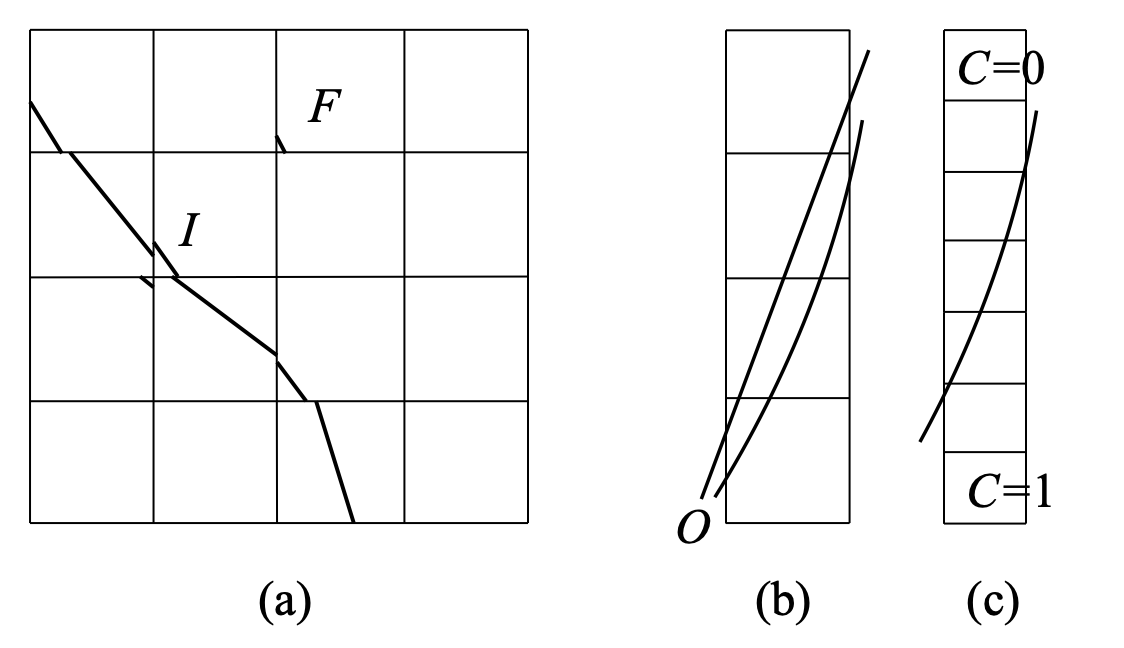}
\end{center}
\caption{(a) A small floatsam $F$ in a cell away from the interface is negligible 
when height functions are used to determine the location of the interface. On 
the other hand requiring the interface to pass near each centroid $C_i$ has a 
large effect. A small inconsistency such as near point $I$ is also ignored by 
the height function. (b) Two cases where the height function expression 
(\ref{hfeq1}) is appropriate with four cells ($n_c=3$, see text). For more 
vertical lines or larger curvatures the line exits the $4 \times 1$ stencil 
through the top and bottom and the HF method cannot be used. (c) To check the 
validity of a the HF calculation one needs to have one full cell ($C=1$) below 
and one empty cell ($C=0$) above, or the converse.}
\label{floatsam}
\end{figure}

We give some details about height functions since it is a relatively novel 
aspect of the code. Height functions were introduced in \cite{popinet09} and 
further discussed, tested and improved in several papers \cite{Bornia11,owkes15}. 
A height function is a function on the discrete grid that gives \red {the elevation 
of the interface with respect to a reference plane}. The use of height functions 
greatly improves the accuracy of VOF methods since it allows us to neglect 
small inconsistencies in the VOF representation. On one hand a small VOF floatsam 
in a cell has only a very small influence on the height-function calculation. 
On the other hand if taken as an indication of the presence of \red {the} interface 
in  the given cell it would create a large error on the interface location 
(\red {see} Fig. \ref{floatsam}a). With high resolution, height functions can 
also yield the position of the interface to fourth-order accuracy \cite{Bornia11}.
We consider for simplicity the case of an approximately horizontal interface, 
\red {then the reference plane is aligned with the $x,y$ plane}. A 
\red {local height} $H$, rescaled by the grid size $h$,  may be defined as 
\be
H = \sum_{p=k_0}^{p=k_0+n_c} C_{i,j,p}. 
\label{hfeq1}
\nd
and is illustrated on Fig. \ref{floatsam}b. \red {The expression \eqref{hfeq1} 
defines a ``vertical'' height with the reference plane passing through point $O$ 
and} the base of the bottom cell on Fig. \ref{floatsam}b.  
More general cases, with arbitrary orientation of the interface, are considered 
in  \ref{app:A}.

\red {The computation of the curvature may then be performed
by  reconstructing a polynomial approximation to the height function}. 
To illustrate this, we take again the case of an approximately horizontal 
interface. Then we fit \red {the local heights by the function} 
\be
H(x,y) = \frac{a_1}2 x^2 +  \frac{a_2}2 y^2 + a_3 xy + a_4 x + a_5 y  + a_6   
\label{apoly}
\nd
where the coefficients \red {$a_i$ are computed from the heights data 
using finite differences (see Eq. (\ref{apolycoef}))}. The curvature is 
then
\be
\kappa=\eps\frac {a_1(1+a_5^2) + a_2(1+a_4^2) - 2a_3a_4a_5}{(1+a_4^2+a_5^2)^{3/2}} 
\label{kpoly}
\nd
where $\eps=1$ if the interface is in the ``canonical'' position (normal pointing 
upwards). The above method is possible only if \red {all nine vertical heights are 
available in the $x,y$ plane}. If they are not, fallback methods are used, details 
of which are given in  \ref{app:A}.

\begin{figure}[!htb]
\begin{center}
\includegraphics[width=0.8\columnwidth,angle=0]{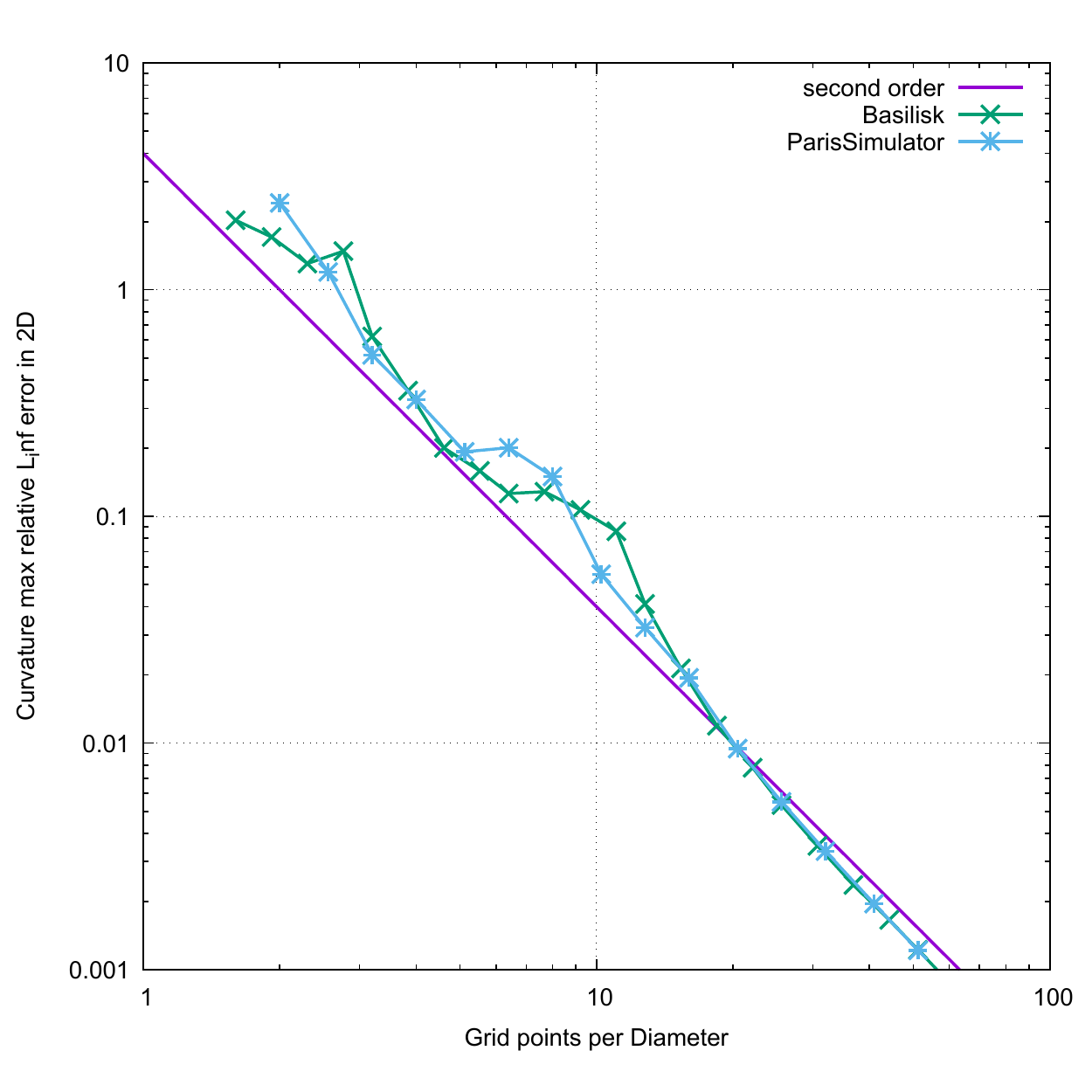}
\end{center}
\caption{Maximum $L_\infty$ error norm in two dimensions for the curvature estimated 
for a cylinder using the height function method in \pariss simulator and \basilisk. 
The mixed-height option is set in both codes.}
\label{owkes2}
\end{figure}

\begin{figure}[!htb]
\begin{center}
\includegraphics[width=0.8\columnwidth,angle=0]{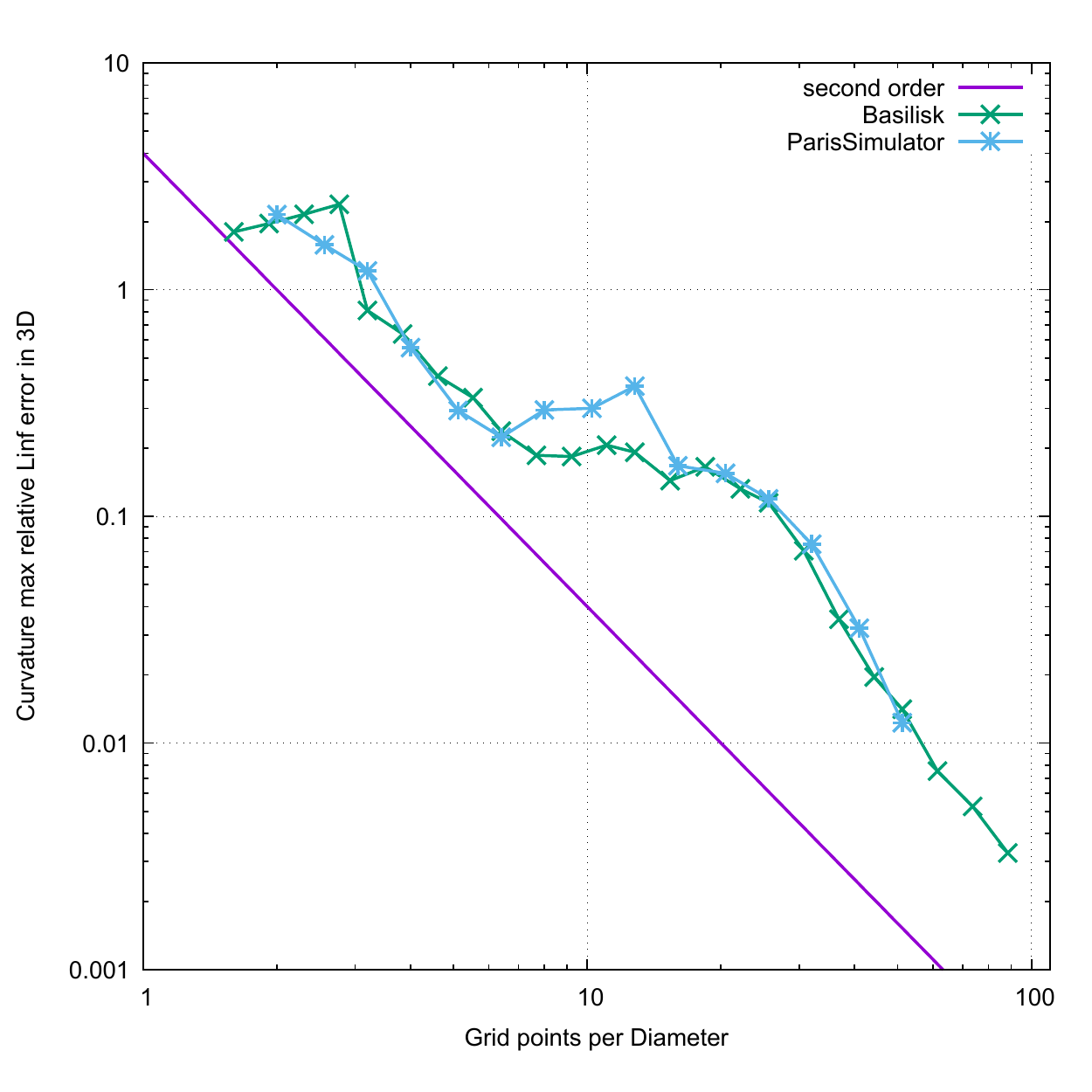}
\end{center}
\caption{Maximum $L_\infty$ error norm in three dimensions for the curvature estimated 
for a sphere using the height function method in \pariss and \basilisk. The mixed-height
option is set in both codes.}
\label{owkes3}
\end{figure}

The performance of our method for the computation of curvature is
shown in Figs. \ref{owkes2} and \ref{owkes3}.  The error was
computed for a collection of diameter-to-cell-size ratios $D/h$. For
each value of $D/h$ the error computation was repeated for an ensemble
of $N$ sphere centers located randomly. The $L_{\infty}$ norm for a given
sphere center is the maximum difference between the sphere curvature
and the numerically obtained curvature. The error reported on \red {these two}
figures is the maximum $L_{\infty}$ error for the whole ensemble of $N$
spheres. We checked that the error varies little when $N$ is increased
above 16, and the standard test case for curvature distributed with
the code uses this value of $N$.  The differences between the error
norms obtained for \basilisks and \pariss are discussed in  \ref{app:A}.

\subsection{Pressure solver} 
\label{pressuresolver}

\subsubsection{In-code Gauss-Seidel solver} 
\label{gss}
The default Poisson solver used to invert the elliptic operators
appearing in Eqs. \eqref{poisson} and \eqref{muelliptic}
is a red-black Gauss-Seidel (GS) solver with overrelaxation \cite{Briggs87}.
\red {Both equations can be discretized in the form}
\begin{eqnarray}
  A_{1,i,j,k} \,p_{i-1,j,k} & + & A_{2,i,j,k}  \,p_{i+1,j,k} + A_{3,i,j,k}  \,p_{i,j-1,k} + \nonumber \\
 A_{4,i,j,k} \,p_{i,j+1,k} & + & A_{5,i,j,k}  \,p_{i,j,k-1} + A_{6,i,j,k}  \,p_{i,j,k+1} - \nonumber \\
 A_{7,i,j,k} \,p_{i,j,k}   & = & A_{8,i,j,k} \,,
\end{eqnarray}
\red {where the coefficients verify the relations}
\begin{eqnarray}
A_{7,i,j,k} & = &  \sum_{p=1}^6  A_{p,i,j,k} \\
A_{2,i,j,k} & = & A_{1,i+1,j,k}\\
A_{4,i,j,k} & = & A_{3,i,j+1,k}\\
A_{6,i,j,k} & = & A_{5,i,j,k+1}\\
\end{eqnarray}
and are constructed by interpolations of $1/\rho$, \red {for Eq. \eqref{poisson},}
or $\mu$, for the \red {implicit part \eqref{muelliptic} of the 
viscous term in the momentum equation}.
The GS solver iterates the assignment
\begin{eqnarray}
  p_{i,j,k} & \leftarrow &   \left(1-\beta \right) p_{i,j,k} +  \frac{\beta}{A_{7,i,j,k}}  \Big(A_{1,i,j,k} \, p_{i-1,j,k}   \nonumber \\
&+& A_{2,i,j,k} \, p_{i+1,j,k} + A_{3,i,j,k} \, p_{i,j-1,k} +  
A_{4,i,j,k} \, p_{i,j+1,k}  \nonumber \\
&+& A_{5,i,j,k} \, p_{i,j,k+1} + A_{6,i,j,k} \, p_{i,j,k-1} -  A_{8,i,j,k}\Big) 
\label{eq:beta_pressure}
\end{eqnarray}
where $\beta$ is an overrelaxation parameter that can be set by the user. 
A value of $\beta=1.3$ is typically used.

\subsubsection{ library multigrid solver}
\label{hls}
The {\sc Hypre} library, developed \red{at} the Lawrence Livermore National Laboratory
(LLNL), is also an option to solve the elliptic equations with multigrid iterative
methods. Since a structured grid is used in \paris, \red {a few solvers of} the {\sc Hypre}
library can be used. The SMG and PFMG multigrid solvers have been implemented in the code 
and used for large-scale simulations using up to 64,536 cores. Both SMG and PFMG
are parallel semicoarsening multigrid solvers. The difference lies in that the SMG
solver uses plane smoothing while the PFMG solver uses pointwise smoothing. The 
plane-smoothing feature makes the SMG solver more robust but less efficient. 
\red {In fact}, the scaling performance of the PFMG solver is much better than the 
SMG solver, since the smoothing \red {operations only involve} a local stencil.

In order to take advantage of the higher efficiency of PFMG and the robustness of SMG,
a solution strategy has been implemented in the code. The PFMG solver is used by
default, \red {however} if the iteration diverges or fails to converge within 
the maximum iteration
number, the code will switch to the SMG solver and redo the iteration. If the
iteration converges, then the code will switch back to PFMG for the next time step.
For a large-scale simulation that runs for a long time, this strategy has been shown
to achieve a good performance, balancing robustness and efficiency.

The {\sc Hypre} library, at least in the versions we use, appears to control the tolerance
on the residual using the $L_1$ norm. The code \red {then} recomputes the residual
norms and controls the accuracy using a norm of the residual chosen by the user 
among $L_1$, $L_2$ and $L_\infinity$.

\subsubsection{In-code multigrid solver} 
\label{mgs}
The code has also a native implementation of a multigrid solver for
structured grids with $2^n$ number of points per \red {coordinate} direction.  
In particular the V-Cycle scheme is implemented and fully parallelized
\cite{Briggs87}.  Relaxation operations are applied starting \red {first} from 
the finest to the coarsest \red {grid}, and then from the coarsest to the
finest one, the number of relaxation \red {operations}  being a user-adjustable
parameter. One advantage of having a native multigrid solver is that
it allows for an efficient solution of the Poisson equation without
the necessity of having external libraries installed in the
system, \red {such as {\sc Hypre}}. Especially when running heavy 
\red {three-dimensional} simulations
in parallel the use of this native solver has been shown advantageous
in some systems with respect to {\sc Hypre} in terms of memory manipulation.

\subsubsection{GPU-accelerated solver}
\label{gpu}
A \red {GPU-based} solver is also available for solving the Poisson equation
when a significant number of iterations is required to achieve
convergence.  The pressure is solved using a Jacobi method for 
Eq. (\ref{eq:beta_pressure}). The need for the Jacobi method
instead of a Gauss-Seidel arises because of the intrinsic nature of GPU
devices. The usual domain decomposition parallelization allows the
implementation of the iterative step by using a simple \textbf{for} loop over the
indexes $i,j,k$. The sequentiality of the indexes cannot be achieved
on GPU devices, as in this case each index combination is ideally
computed simultaneously. In this sense, the larger number of iterations
required by the Jacobi method is mitigated by the speed-up provided by
GPUs.

The memory handling is a critical aspect in GPUs applications and it
is even more critical in DNS. Although a Jacobi method intrinsically
requires doubling the memory usage for the \red {pressure} matrix $p$, 
it also enables the leanest data transfer between CPUs and GPUs, which 
is a critical aspect of normal CUDA applications. A Gauss-Seidel
red-black solver is in principle possible and \red {could be} beneficial 
for certain applications. In fact, let us assume that the size of the 
matrix $p$ is $N_p=n_x n_y n_z$, the normal implementation of a red-black 
Gauss-Seidel solver would be
\begin{algorithmic}

\FOR{all $\Omega_{i,j,k}$ cells of "red" type} 
\STATE{compute $p_{i,j,k}$ using (\ref{eq:beta_pressure}) }  
\ENDFOR

\FOR{all  $\Omega_{i,j,k}$  cells of "black" type} 
\STATE{compute $p_{i,j,k}$ using (\ref{eq:beta_pressure}) }  
\ENDFOR

\STATE{check convergence}
\end{algorithmic}
which inherently reduces the memory usage required. On the other hand,
the first \textbf{for} loop is not parallelizable in an efficient way
in \textit{CUDA}. Therefore, such an algorithm would be beneficial
from a memory standpoint, but would improve the computational time
only if $N_p$ is at least 4 times greater than the number of GPU
process available. As \red{this} is usually not the case, the beneficial
effects of a red-black algorithm are limited, although it will be
object of future studies.

The implementation of the algorithm is achieved by means of the
open-source \textit{CUDA} library for \textit{C} developed by NVIDIA,
while the intercommunication between processors is still achieved by
using MPI. For this reason, an interface between \textit{Fortran90}
and \textit{C} is created in {\sf module\_CUDA.f90}. By passing
through the interface, each process transfers the \red {coefficients 
matrix} $A$ and \red {the pressure matrix} $p$
to the \textit{C/CUDA} environment (in {\sf poissonCUDA.cu}) where the
iteration step is performed. The boundary conditions, as well as the
MPI communication, are enforced in the environment that originally
created the MPI communication, hence these functions are programmed in
{\sf cudaFun.f90}.

\subsubsection{\red {Free-surface flow solver}}
\label{fsps}
A free-surface flow solver is implemented in \paris, which is designed to apply a 
free-surface condition as described in Section \ref{subsec:fs}.
\red {The implementation of the boundary conditions 
(\ref{free_surf_norm_bc},\ref{free_surf_kine_stress}) is here limited to inviscid 
flows, hence only the jump
condition \eqref{free_surf_norm_bc} along the normal direction is considered.}
The free-surface solver uses the VOF method \red {implemented} in \pariss to track the
interface. The flow is then solved \red {only in the phase with $\Heaviside=0$}, using 
the same numerical methods \red {described previously}. For the purpose of description 
the solved phase will be called liquid and the unsolved phase will be gas. 
\red {In the region occupied by the gas phase the conservation equations are not 
solved, instead a homogeneous pressure field is assumed, that can vary only in time.}
The time variation \red {of the pressure is determined by} a polytropic gas law
\begin{equation}\label{eq:polygas}
p_c = p_0 \left(\frac{V_0}{V_c}\right)^{\gamma} \,,
\end{equation}
where $V_{c}$ is the total volume of \red {the} gas pocket at pressure $p_c$, $p_{0}$ 
and $V_{0}$ are respectively the reference pressure and volume of the gas phase, 
\red {and finally} $\gamma$ is the heat capacity ratio. 
The gas phase pressure along with the pressure jump due to surface tension is then 
applied as a Dirichlet boundary condition for the pressure 
in the liquid flow, as given in Eq. \eqref{free_surf_norm_bc}.

\begin{figure} 
  \centering
  \def\svgwidth{0.55\columnwidth}
  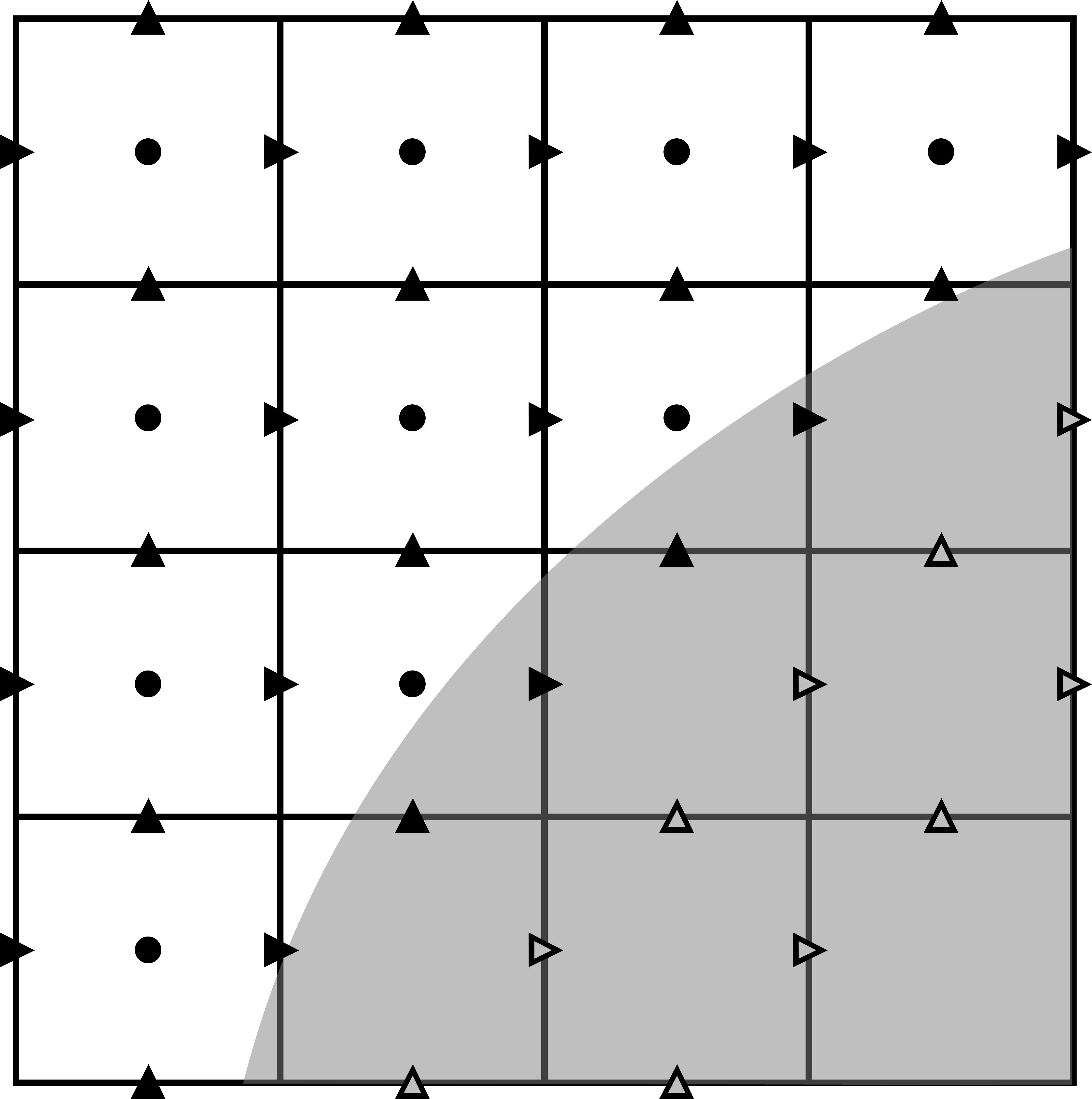
  \caption{A 2D section of the numerical grid, showing part of a gas bubble in grey. 
  Circles represent computational cell nodes where pressure is calculated. Triangles 
  indicate scalar velocity components on the computational cell faces. Filled triangles
  indicate values which are found by solving the governing equations, while unfilled 
  triangles represent boundary values found by extrapolation.}
  \label{fig:p_nodes}
\end{figure}

The method used to apply this pressure boundary condition is inspired by the idea of 
Fedkiw and Kang \cite{Fedkiw99,Kang00}, often referred to as the \red {Ghost Fluid 
Method (GFM)}. Let the time-varying pressure in the unsolved phase be $p_c$. 
Special care is required in the discretization of \red {the elliptic equation} 
\eqref{poisson} for liquid cells near the interface. Cells that 
contain mostly gas are excluded from the solution, so that only cells where $C<0.5$ are 
solved (\red {we recall that} the convention is to have $C=0$ in the liquid). Figure \ref{fig:p_nodes} shows a representation of a 2D grid with a section of an interface. 
The grey area represents a gas-filled volume. Cells that contain a filled circle are 
included in the pressure solution, while cells without a marker \red {in the center} 
are excluded.
Since only the liquid phase is solved, only the liquid density is applied. Furthermore, 
for cubic cells $\Delta x = \Delta y =\Delta z = h$, where $h$ is the constant grid 
spacing.

\begin{figure}
  \centering
    \def\svgwidth{0.55\columnwidth}
    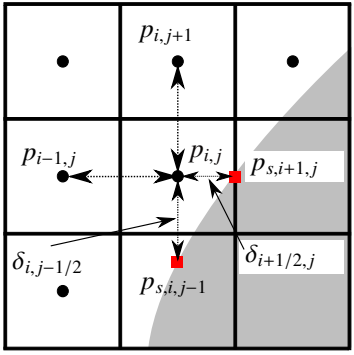
    \caption{Discretisation of the pressure equation near the interface}
  	\label{fig:p_mod_branches}
\end{figure}
The stencil for the pressure gradient components has to be changed near the interface when 
a neighbouring cell falls inside the gas phase. The pressure in this cell must be 
substituted by a surface pressure $p_s$. We apply a \red {finite-difference} gradient
approach as Chan \cite{Chan70}. As an example, the approximation for the pressure 
gradient components for the cell with indexes $i$ and $j$ in Fig.
\ref{fig:p_mod_branches} is written
\begin{equation} \label{eq:mod_branches}
\nabla^{h}_x \,p_{i+\rfrac{1}{2},j} = \dfrac{p_{s,i+1,j} -
p_{i,j}}{\delta_{i+\rfrac{1}{2},j}} \,\,;\quad
{\nabla}^{h}_y \,p_{i,j-\rfrac{1}{2}} = \dfrac{p_{i,j} -
p_{s,i,j-1}}{\delta_{i,j-\rfrac{1}{2}}} \, ,
\end{equation}
where $\delta$ is the distance between the pressure node under consideration and the
intersection with the interface. The \red {surface pressure} $p_{s}$ on the liquid side 
of the interface is found by adding to \red {the gas pressure} $p_c$ the Laplace pressure
jump. The pressure $p_c$ inside each cavity is known from the polytropic law \eqref{eq:polygas}. The interface pressure \red {$p_s$ in the first expression of
\eqref{eq:mod_branches}} will then be
\begin{equation}
p_{s,i+1,j} = p_{c,i+1,j} + \sigma \frac{\kappa_{i,j} + \kappa_{i+1,j}}{2} \,.
\label{Laplace_num}
\end{equation}
From Eqs. \eqref{eq:mod_branches} and \eqref{Laplace_num} it is clear that
\red {an accurate computation of the} interface curvature as well as an accurate 
prediction of the interface location are important parameters to ensure the accuracy 
of the pressure solution. Since the height function is the approximate interface 
distance from some reference cell \red {boundary} in a given direction, it is used for 
$\delta$. When the interface configuration is such that a height cannot be obtained 
in the required direction, the distance is approximated by using a plane reconstruction 
of the interface in the staggered volume. 

%
%
\paragraph{Extrapolation of the velocity field} 
\label{sec:velocity_extrapolation}

The resolved velocity components right next to the interface will require neighbours in the
gas phase to discretize the momentum advection term. These values in the gas phase can be 
seen as boundary values to the resolved velocities. In order to find neighbours in the gas
phase, we extrapolate the resolved velocities similarly to Popinet \cite{popinet02}.

After calculating the liquid velocities using standard methods in \paris, the boundary velocities in the gas phase are updated for the next time step from the closest two velocity neighbours using a linear \red {least-square} fit. Let \red {us assume that} the velocity
field can be described as a linear combination
\begin{equation}
\U \left(\X\right) = \AA
\cdot (\X - \X_{0} ) + \U_{0}
\end{equation}
where the components of the tensor $\AA$ and of the vector $\U_0$ are the unknowns.
\red {In two dimensions we take} a $5 \times 5$ stencil around the unknown gas velocity 
at location $\X_0$ \red {and} find the extrapolated velocity $\U_0$ by minimizing the
functional
\begin{equation}
\mathcal{L} = \sum_{k=1}^{N} \big | \AA \cdot
(\X - \X_0) + \U_0 -
\U_k \big |^{2}
\end{equation}

This is done first for all locations closest to the resolved velocities $\U_k$ (``first
neighbours''), whereafter the process is repeated for the ``second neighbours''. Note that
only resolved velocity components are included in the cost function, therefore the number 
of \red {the known} velocities $N$ can vary depending on the shape of the interface.
Furthermore, because of the staggered grid, only one \red {component of the velocity} 
$\U_0$ is computed at any location $\X_0$.

\paragraph{Ensuring volume conservation}\label{sec:2nd_projection}

An additional step is required to ensure that the extrapolated velocities are \red
{divergence-free}. This is required to ensure that the advection of $C$ is conservative.

\begin{figure} 
  \centering
  \def\svgwidth{0.55\columnwidth}
  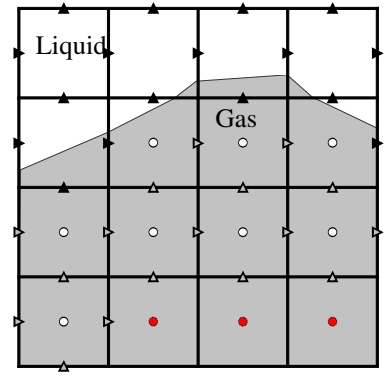
  \caption{2D example of the problem to correct the extrapolated velocities (unfilled
  triangles). A \red {Poisson's} problem is solved in the cells marked with an unfilled
  circle.}
  \label{fig:2nd_proj}
\end{figure}

A similar approach to Sussman \cite{Sussman03} is used. Only the first two layers 
of cells inside the gas phase are considered and all other cells are disregarded. 
A 2D example is presented in Fig. \ref{fig:2nd_proj}.
Similar to the projection step explained earlier, a ``phantom'' pressure is obtained in
these cells by solving a \red {Poisson's} equation 
\begin{equation}
\nabla^{h} {\cdot} \bigg ( \nabla^{h} \hat{P} \bigg ) = 
\nabla^{h} {\cdot} {\tilde{\U}} \,,
\end{equation}
where $\hat{P}$ is the ``phantom'' pressure and ${\tilde{\U}}$ is the \red {extrapolated}
velocity on the faces of the first two gas neighbours. $\hat{P}$ is only calculated in
the cells represented by unfilled \red {circles} in Fig. \ref{fig:2nd_proj}. On the
liquid side of these cells, the solved velocities (filled triangles) are used as a
velocity boundary condition with the pressure gradient on this face set to zero. On the
gas side, \red {a fixed pressure is prescribed in the cells (red filled circles) outside
the region where the ``phantom'' pressure is computed.}
Only the extrapolated velocities (unfilled triangles) are then corrected by the solved
pressure gradient $\nabla \hat{P}$ 
\begin{equation}
{\tilde{\U}}^{n+1} = {\tilde{\U}} - \nabla^{h} \hat{P}
\end{equation}
to ensure a \red {divergence-free velocity field} in the first two layers of cells just 
inside the gas.

For more details on the numerical method and its application in idealized 
\red {\em microspalling}, see \cite{anisz18} and \cite{malan19}. \red {This process is 
found in metals that under shock loading melt and where, with the reflection of the 
shock wave from the material free surface, cavities can nucleate, grow and merge.}

\subsection{Solid boundaries}
\label{solids}
Solids are defined in a ``block'' or ``Lego'' manner. A domain-wide binary flag
$s_{i,j,k}$ is defined that is equal to 0 inside the solid and 1
outside. A set of link-based flags $s^x$, $s^y$ and $s^z$ is also
defined (a link-based array is data located on the velocity component
collocation points, such as $i+1/2,j,k$ \red {for the horizontal velocity
component $u$}). The following pseudo code is executed at initialisation
\red {relatively to the $x$ direction, with $i=1,2,...,n_x$

\begin{algorithmic}
\FOR{all $j,k$} 
\STATE{$s^x_{1/2,j,k} \leftarrow s_{1,j,k}$}
\ENDFOR
\FOR{all $i,j,k$} 
\STATE {$s^x_{i+1/2,j,k} \leftarrow s_{i,j,k}$ } 
\ENDFOR
\end{algorithmic}
and similarly for the $y$ and $z$ directions}.
The indexes $s^x$, $s^y$ and $s^z$ are then used to ``block'' the velocity and the 
pressure correction on the solid region and its boundary. This is done each time the 
velocity is updated

\begin{algorithmic}
\FOR{all $i,j,k$} 
\STATE{$u_{i+1/2,j,k} \leftarrow \red {s^x_{i+1/2,j,k}} \, u_{i+1/2,j,k} $}  
\ENDFOR
\end{algorithmic}
\red {and similarly for the velocity components $v$ and $w$.}
The pressure correction should not change the velocity \red{on the solid boundaries}, 
so on the links a Neuman boundary condition for the pressure is established, which is
equivalent to setting to zero some coefficients
\begin{algorithmic}
\FOR{all $i,j,k$} 
\STATE{$A_{1,i,j,k}   \leftarrow s^x_{i-1/2,j,k} \, A_{1,i,j,k}$}  
\STATE{$A_{2,i,j,k}   \leftarrow s^x_{i+1/2,j,k} \, A_{2,i,j,k} $}
\STATE{(similarly for $A_3,A_4$ and $s^y$, and for $A_5, A_6$ and $s^z$)}
\STATE{$A_{7,i,j,k} \leftarrow  \sum_{p=1}^6  A_{p,i,j,k}$}
\ENDFOR
\end{algorithmic}
The solid domain can be initialized by implicit functions or by
loading a file containing the $s_{i,j,k}$ data. In both cases it is
important that the presence of the solid does not make the linear system
(\ref{poisson}) ill-posed. This will happen for example if the boundary
conditions are Dirichlet for the velocity at the entry of a channel
and the solid completely blocks the channel. \red {To avoid this type of
problem,} there is a small utility program ``{\sf rockread.c}'', 
distributed with the code, that checks that the fluid ``percolates''.

\subsection{\red {Lagrangian point-particle model}}
\label{lpp}

\red{
A Lagrangian point-particle (LPP) model has been implemented in PARIS, that is fully 
coupled to the VOF method to provide a multi-scale modeling strategy to simulate liquid
atomization: the large-scale interfaces on the bulk liquid are resolved by VOF, while 
the small droplets are represented by the LPP model. The details of the model can be 
found in \cite{ling15} and only a brief summary is given here. 

\subsubsection{Overview}
The combined VOF-LPP algorithm is shown in Fig.\ \ref{fig:LPP_flowchart}. 
The Navier-Stokes equations and the advection equation for the volume fraction
$C$ are solved for the resolved flow. Then the $C$ field is tagged to identify 
different liquid structures. The tagging approach of \cite{herrmann10} is used, 
and cells that are attached to each other, with respect to the liquid phase, will have 
the same tag number.
The size, aspect ratio, and distance from the bulk flow of each liquid structure are
computed in the tagging process. Small resolved droplets (RD) that satisfy the LPP 
criteria will be submitted to the RD-to-LPP conversion routine.
The motion equation is then solved for each LPP droplet to update its velocity and
position. Afterwards, the droplet is examined to decide whether or not
it should be converted back to a resolved droplet.  

LPP droplets and the resolved flow (RF) are two-way coupled. The RF properties, such 
as fluid velocity, are needed to calculate the forces acting on each LPP droplet. 
On the other hand, these forces need to be subtracted from the momentum equation 
of the flow, thus appearing as an additional source term. 

\begin{figure}[tbp]
  \center
  \includegraphics[width=0.85\columnwidth]{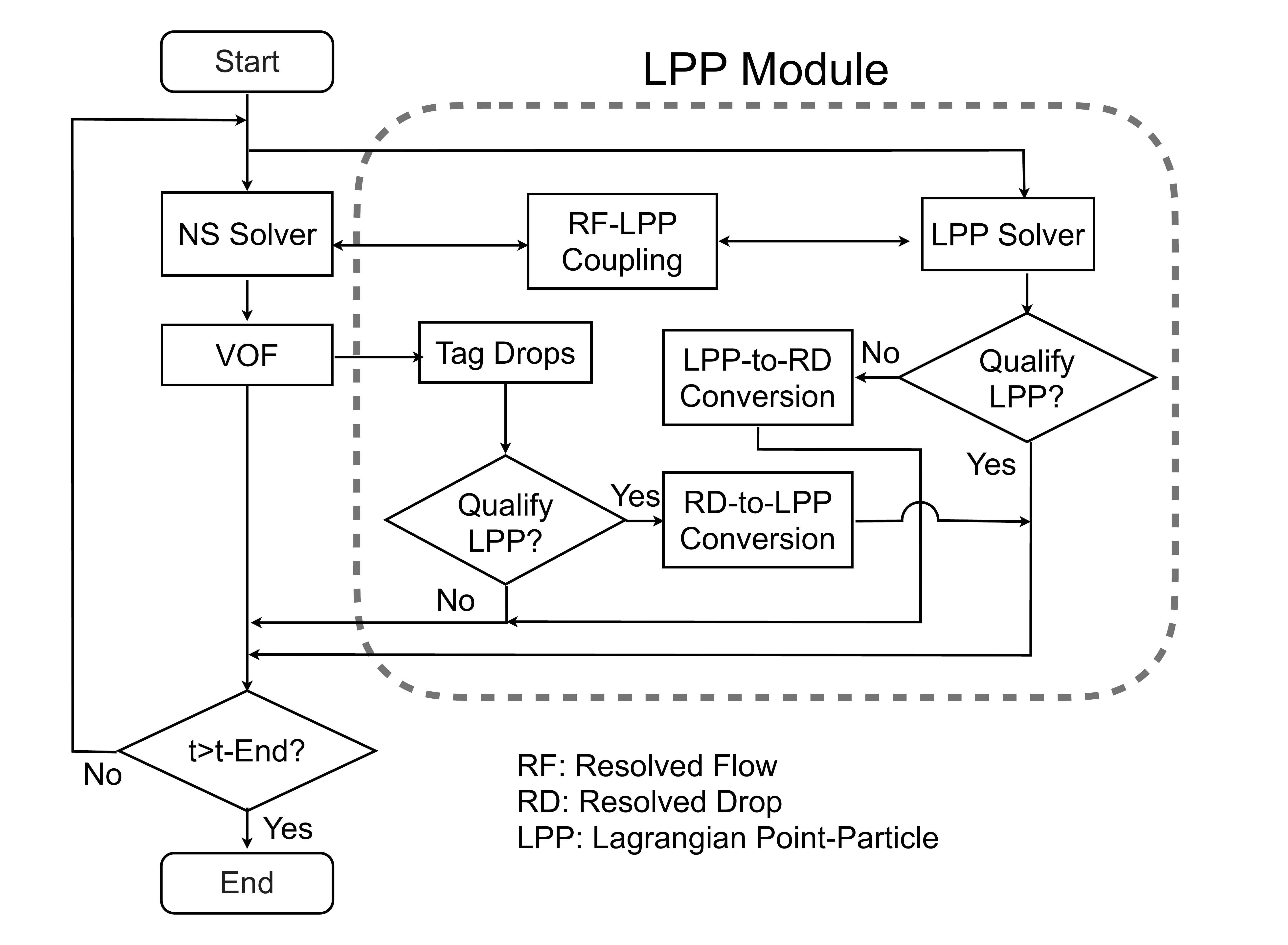}
  \caption{\red {Flow chart of the combined VOF-LPP algorithm.}}
  \label{fig:LPP_flowchart}
\end{figure}

\subsubsection{Two-way coupling between LPP and RF}
The non-conservative momentum equation (\ref{nsnoncons},\ref{nsnonconsbis})
is now written as
\be
\rho \big( \dert \U \,+\, \U \cdot \nabla \U \big) = - \grad p + \grad \cdot \DD +
{\sigma \,\kappa \,\delta_S \,\N} - \F_p
\label{eq:mom}
\nd
%
where $\F_p$ is the closure term that accounts for the backward effect of the LPP 
droplets on the resolved flow. 
The LPP model approximates each small droplet as a point mass, hence the droplet-scale 
flow is not resolved. To accurately track the LPP droplets in the Lagrangian framework, 
the force exerted on each droplet is calculated in terms of the undisturbed flow field
properties. The closure is typically given 
by the force model or the so-called equation of motion (EOM)
\cite{crowe98}
\begin{align}
\frac{d \X_p}{d t} & =  \U_p\, ,
\label{eq:part_pos} \\
\frac{d \U_p}{d t} & =  \frac{{\U} - \U_p}{\tau_p} \phi
+ \frac{\rho}{\rho_p}  \frac{D {\U}}{D t} 
+ \frac{C_m \rho}{\rho_p} \left( \frac{D {\U}}{D t} 
- \frac{d \U_p}{d t} \right) + \G \, , 
\label{eq:part_eom}
\end{align}
where $\rho$ and ${\U}$ are the density and velocity of the undisturbed flow,
and $\X_p$, $\U_p$, $\tau_p={ \rho_p \,d_p^2}/(18 \,\mu)$, $\rho_p$, and $C_m$ 
the position, velocity, response time, density, and added-mass coefficient of 
the LPP droplet. A tri-linear interpolation is used to interpolate 
the fluid velocity from the grid to the position of the LPP droplets.
The modified gravity acceleration is denoted by 
$\G = (1-\rho/\rho_p)\G'$, where $\G'$ is the gravity acceleration.
For the time integration of Eqs. \eqref{eq:part_pos} and \eqref{eq:part_eom}
a second-order predictor-corrector method is used, which is consistent with the 
algorithms used for the resolved flow. 

The force terms on the right hand side of Eq. \eqref{eq:part_eom} represent the 
quasi-steady force, the pressure-gradient force, the added-mass force, and the gravity 
force, respectively. The Basset-history force has been ignored for simplicity,
while the Fax\'en and lift forces have been neglected because the LPP model is here 
applied to small droplets, and the effect of inhomogeneous ambient flows on the 
droplet force is expected to be small. 
The quasi-steady force is expressed as the Stokes drag multiplied by a correction 
term $\phi$, which is a function of the Reynolds number 
$\mathrm{Re}_p = {\rho \,d_p \,|{\U}-\U_p|}/{\mu}$ \cite{clift70},
\begin{align}
\phi(\mathrm{Re}_p) &= 1 + 0.15 \,\mathrm{Re}_p^{0.687} + 0.0175 \,\mathrm{Re}_p 
\left( 1 + \frac{42500}{\mathrm{Re}_p^{1.16}} \right)^{-1}\, .
\label{eq:cor_func_Re} 
\end{align}
In atomization, the inter-droplet interaction can be ignored, as the
small droplets are quickly dispersed away from the bulk liquid and the related 
volume fraction is relatively small. 

As a consequence of Newton's third law, the force $\F_p$ exerted on the LPP droplets 
needs to be subtracted from the resolved flow and is referred to as backward coupling. 
This force in the momentum equation \eqref{eq:mom} is calculated as
\begin{align}
 \F_p = \sum_{i=1}^{N_p} \mathbf{F}_{fp,i} \,G(\X-\X_{p,i})\, ,
 \label{eq:force_backeffect}
\end{align}
where $N_p$ is the total number of LPP droplets.
The Gaussian function $G(\X-\X_{p,i})$ is a numerical representation 
of the LPP droplet coupling force \cite{maxey97}
\begin{align}
 G(\X) = (2\pi \mathcal{L}^2)^{-3/2} \exp(-|\X|^2/2 \mathcal{L}^2 ) \, ,
 \label{eq:Gaussian_func}
\end{align}
where $\mathcal{L}$ controls the size of the region where the force 
should be distributed. Note that only the force due to fluid-LPP-droplet 
interaction needs to be subtracted from the momentum equation
\begin{align}
 \mathbf{F}_{fp,i}=m_{p,i} \left( \frac{d \U_{p,i}}{d t}- \G \right) \, .
 \label{eq:coupling_force}
\end{align}

\subsubsection{Two-way conversion between LPP and resolved droplets}
\label{sec:conversion}
The droplets that are generated in the atomization process are converted to LPP 
droplets when they satisfy the following criteria. 

\paragraph{LPP Criteria} 
The criteria to determine whether a resolved droplet (RD) should be represented as 
a LPP droplet are based on its volume $V_p$, aspect ratio $\gamma_p$, and position 
$\X_p$. $V_p$ is required to be smaller than a critical value 
$V_{crit} \simeq (4\,h)^3$, and $\gamma_p$ close to one.
Furthermore, since the current LPP model does not include either droplet 
formation or droplet-interface interaction, only droplets that 
are at a given distance away from the liquid jet interface will be converted. 
The distance is typically chosen to be the droplet diameter $d_p$. The overall 
conversion criteria can then be expressed as
\begin{align}
 \{\mathrm{LPP\ Qualified}\} = \{V_p < V_{crit}\} \,\,\&\&\,\,
 \{|\gamma_p - 1| < \epsilon_{\gamma}\}\,\,\&\&\,\,\{\X_p\in \mathcal{R}_{ai} \} 
 \,,
\end{align}
where $\epsilon_{\gamma}$ is the tolerance for the aspect ratio,
and $\mathcal{R}_{ai}$ is the region away from the interface.

\paragraph{RD-to-LPP conversion}
If a RD satisfies the LPP criteria, the volume fraction $C$ in the corresponding cells 
is set to zero. A new LPP droplet is created and added to the LPP array, together with 
its volume and velocity. Furthermore, the flow field in the same region will be replaced 
by the undisturbed flow, so that the LPP droplet sees the proper undisturbed velocity
field. This is done by interpolating each component of the fluid velocity along the corresponding coordinate from the surface into the interior of a cubic region centered 
at the droplet location. The reconstructed velocity field is globally divergence-free if 
the velocity on the surface of the interpolation region is divergence-free. 
Unless the particle Reynolds number is very small, a cubic region with an edge size twice  
the droplet diameter is sufficient to reconstruct the undisturbed flow field in a 
satisfactory way. 

\paragraph{LPP-to-RD conversion}
When a LPP droplet does not satisfy the LPP criteria anymore, for example when it is
too close to the liquid-jet interface, it will be converted back to a resolved droplet. 
First, a spherical droplet is rebuilt around $\X_p$, by specifying the volume 
fraction $C$ in the cells that will be occupied by the resolved droplet. The velocity 
field needs to be updated in the same cells. To account for the momentum of both the 
droplet and the virtual fluid moving with the droplet, the velocity in the cells 
occupied by the resolved droplet should be changed to $\U'_p$, which is calculated as 
\begin{equation}
  \U'_p - \U = \Big( 1 + \alpha \,\frac{\rho}{\rho_p} \Big)\, 
  (\U_p - \U)\, ,
\end{equation} 
where $\alpha$ is the ratio between the volumes of the virtual fluid and the droplet. 
For the added-mass effect, it is considered that the ratio between the volumes of 
the virtual fluid to be accelerated through the inviscid mechanism and the particle 
is the added-mass coefficient $C_M$, which is equal to 0.5 for a sphere in 
incompressible flows. Due to finite viscous diffusion time scale, 
the history force usually needs to be expressed in integral form. 
Nevertheless, it has been shown in \cite{ling13} that, 
if the particle and ambient fluid acceleration time scales are much larger than 
the viscous diffusion time scale, the history force can also be expressed 
as a non-integral form like the added-mass force, and a viscous-unsteady 
coefficient $C_{vu}$ similar to the added-mass coefficient $C_M$ 
can be derived as
\begin{align}
C_{vu}  \approx 8.51 \left( \frac{0.75+0.105{\mathrm{Re}_p}}{{\mathrm{Re}_p}}
\right)\, .
\label{eq:Cvu}
\end{align}
The viscous-unsteady coefficient $C_{vu}$, can be considered as 
the ratio between the volumes of the virtual fluid to be accelerated 
through the viscous mechanism and the particle. 
The excess momentum added through $\U'_p$ is to mimic 
the effects of the added-mass and history forces on accelerating 
the surrounding fluid around the droplet. 
Therefore, $\alpha$ can be estimated as the sum of $C_M$ and $C_{vu}$
and thus depends on the droplet Reynolds number as well. 
}

\section{Testing}
\label{testing}
The testing of the code is performed automatically. The short version of testing
is done by typing {\sf make test}, the long version by typing {\sf make longtest}.
The short version takes approximately 5 minutes on a laptop with an i7 processor
and the long version takes approximately 25 minutes. All the resulting tests
give a report of ``\green{PASS}'' or ``{\color{red}{FAIL}}''.
Each test is contained in a subdirectory of the {\sf Test} directory. 
The subdirectory corresponding to a test has a self-explanatory name,
e.g. {\sf PresPoiseuille} for the pressure-driven Poiseuille flow.

The tests can be divided into two categories: elementary tests which are basically
sanity checks verifying that the code is not corrupted and finds elementary flows
easily, and more complex test that are in some cases a verification of the code,
comparing it to analytical solutions. However, the verification has not been
pushed very far, since the code is an assembly of methods that have been tested
extensively elsewhere, see for example TSZ for a review of these tests. The more
recent methods such as the ``\red {mass-momentum consistent}'' option for velocity
advection, has been tested extensively in \cite{fuster2018momentum} although 
several test cases of the method are included in the test suite and will be described below. 
\red {In order to avoid
the introduction of more equations we assume a constant surface tension
coefficient $\sigma$ in all the testing section.

In many of the tests the solution computed during the test run is compared to a reference solution.
In some cases the reference solution was computed by the authors at a different resolution and
is included with the code distribution. In other cases the ``reference'' is an near-identical numerical solution performed by the authors and included in the code distribution. In that case the reference should be identical to the solution except in some cases where tiny changes in the code or the implementation create moderately large differences. This is the case for the Raindrop test below.}

\subsection{Elementary tests}

\subsubsection{Poiseuille flow}

An elementary Poiseuille flow \cite{kundu2012fluid} is tested. The
simulation is set up in a $8\times8\times2$ grid in which the system
reduces to a 2D planar flow in the box $(0,1)\times(0,1)$.  The
parameters are $\| \nabla p \|=\mu=\rho=1$.  The boundary conditions
set pressure on the left and right. 
The simulation is continued until the flow becomes \red {stationary}, which 
happens with $10^{-3}$ accuracy around time $1$. This time is reached 
in $1000$ time steps. The explicit version of viscous diffusion is used.

\begin{figure}[!htb]
\begin{center}
\includegraphics[width=\columnwidth]{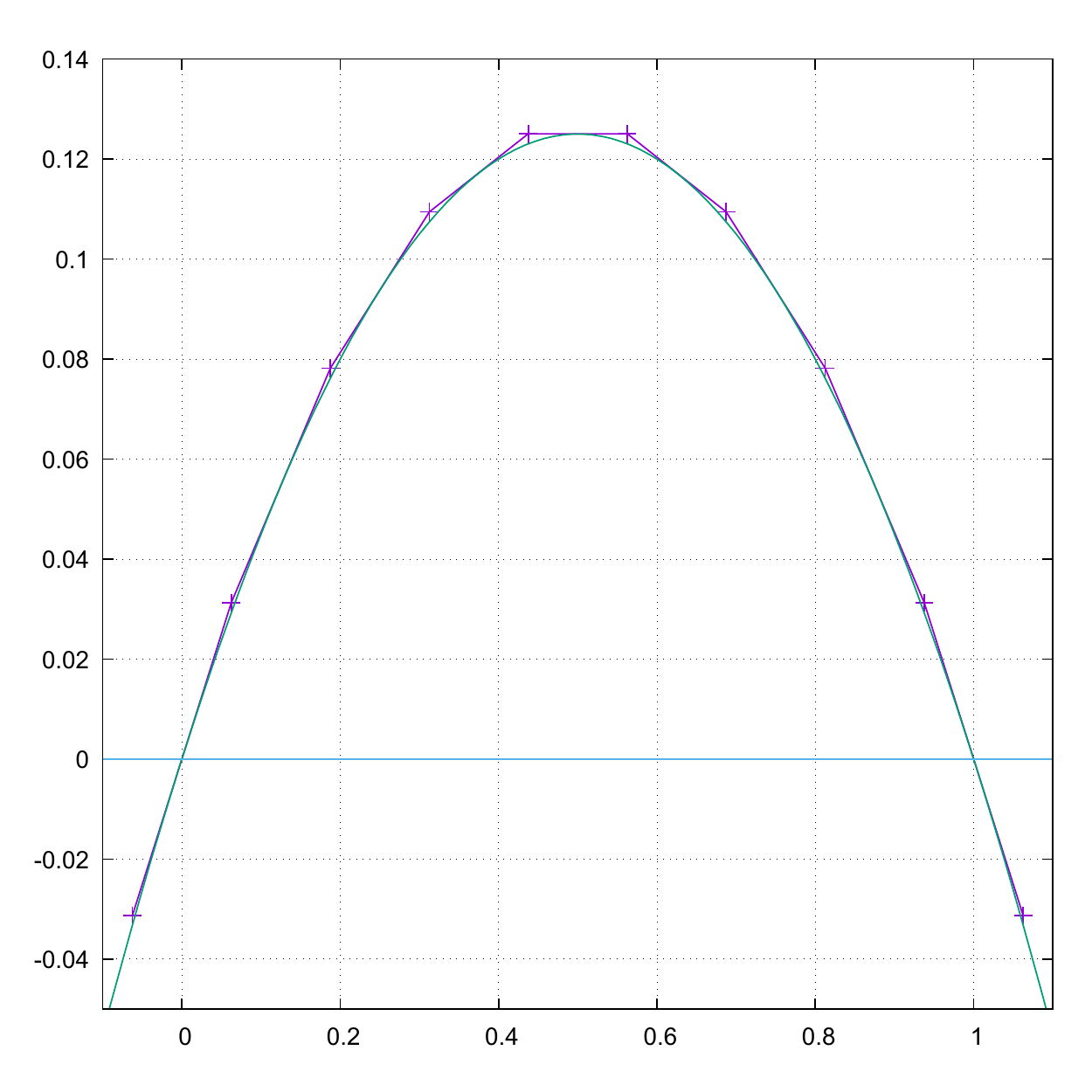}
\end{center}
\caption{Poiseuille flow test case.}
\label{pois}
\end{figure}

This test passes if the
numerical segments are tangent to the theoretical profile as seen in
Fig. \ref{pois}. Because the Poiseuille flow profile is quadratic,
second-order finite differences offer exact values for the second
derivative of velocity, which ensures that the profile found is
exactly a parabola. The accuracy of the \red {parabola amplitude} is
set by the quality of the approximation of the $u=0$
\red {boundary condition (solid wall)}. Since \red {this condition} is set at 
first order, there is a small $\Order(h^2)$ difference with the exact parabola.

The bottom wall tangential velocity  boundary condition is $u=0$ on $y=0$.
Since the wall is at $y_{i,1/2}$, the boundary condition is written using a 
ghost velocity at $y_{i,-1/2}$ that satisfies 
$u^{\rm ghost}_{i,-1/2} + u_{i,1/2}=2 u^{\rm wall}$.
The boundary condition is thus implemented by writing in a ``ghost layer'' 
of the grid
\be
 u^{\rm ghost}_{i,-1/2} \leftarrow 2 u^{\rm wall} -  u_{i,1/2}.
\nd
The result is shown on Fig. \ref{pois}.
It is also possible to run the Poiseuille flow test in other manners, \red {for example}
with set inflow velocity, \red {or} to run it with the implicit version 
of viscous diffusion, in which case the flow converges in a few time steps.

\subsubsection{Stokes flow around a disk}

\begin{figure}[!htb]
\begin{center}
\includegraphics[width=0.9\columnwidth,angle=0]{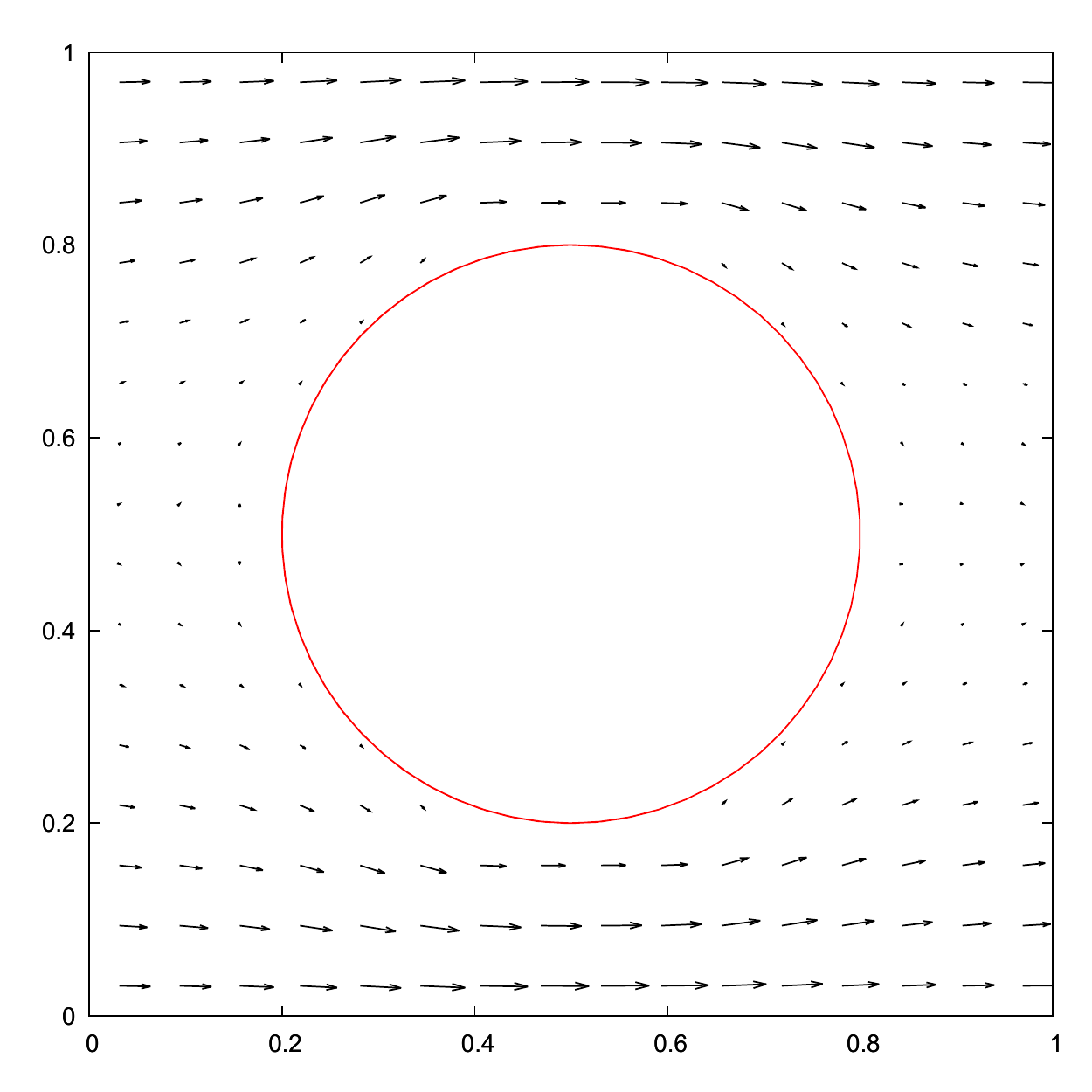}
\end{center}
\caption{Stokes flow around a disk test case.}
\label{disk}
\end{figure}

A pressure driven flow around a disk of diameter $1/2$, with the other parameters 
as before, is set and left to evolve until steady state. The advection operator 
$\LLL_{\rm adv}$ is turned off which ensures that the steady state is a Stokes flow. 
The explicit version of the viscous terms is used. The test resides in the test 
directory {\sf PresDisk}. If the implicit version is used, convergence to the steady
state can be faster but an error of order $\tau$ affects the solution.
The result is shown on Fig. \ref{disk}.
There are two other similar tests beyond this first one. In the second one, located 
in the directory {\sf Disk},  the flow is driven by a body force akin to gravity 
instead of being driven by pressure. This second test is still with periodic 
boundary conditions. The third test has inflow and outflow conditions and is located 
in the test directory {\sf Inflowdisk}. 

\subsubsection{Droplet advection}

This is an elementary test to check whether the VOF method is operating normally. 
The final state of the \red {volume fraction} field $C$ is compared to a precomputed
value. The test has to be visualized ``by hand'' by the user, with the help of 
graphics software such as {\sc Visit} or {\sc Paraview}. One should see an 
\red {undeformed} droplet \red {moving across} the domain. 

\subsubsection{Cylinder advection}

This is a more sophisticated test that probes the ``\red {mass-momentum consistent}''
option. The test is described in detail in \cite{fuster2018momentum}. If the velocity
field stays uniform and constant and as a result the droplet is advected undeformed, 
it means that momentum $\rho \U$ and density $\rho$  are advected in lock-step, so 
that the operation $\rho \U / \rho$, at each time step, gives the constant $\U$. 

\subsubsection{Speed}

This is not so much a test as a report on the code speed. On a 2015 MacBook, a single processor 
run delivered a speed of $1.6\, 10^6$ cells per second. On an AMD EPYC 7351,  a single processor 
run delivered a speed of $1.9\, 10^6$ cells per second and a parallel eight-processor run delivers a speed
of  $1.65\, 10^6$ cells per second and per processor (The parallel test is optional and can be run by the user by typing
{\sf sh run-speed-test.sh $N$} where $np = N^3$ is the number of processors desired).  A significant drop 
of the code speed from this value would be a serious issue.

\subsection{Capillary wave}
\begin{figure}
\begin{center}
\includegraphics[width=0.9\columnwidth,angle=0]{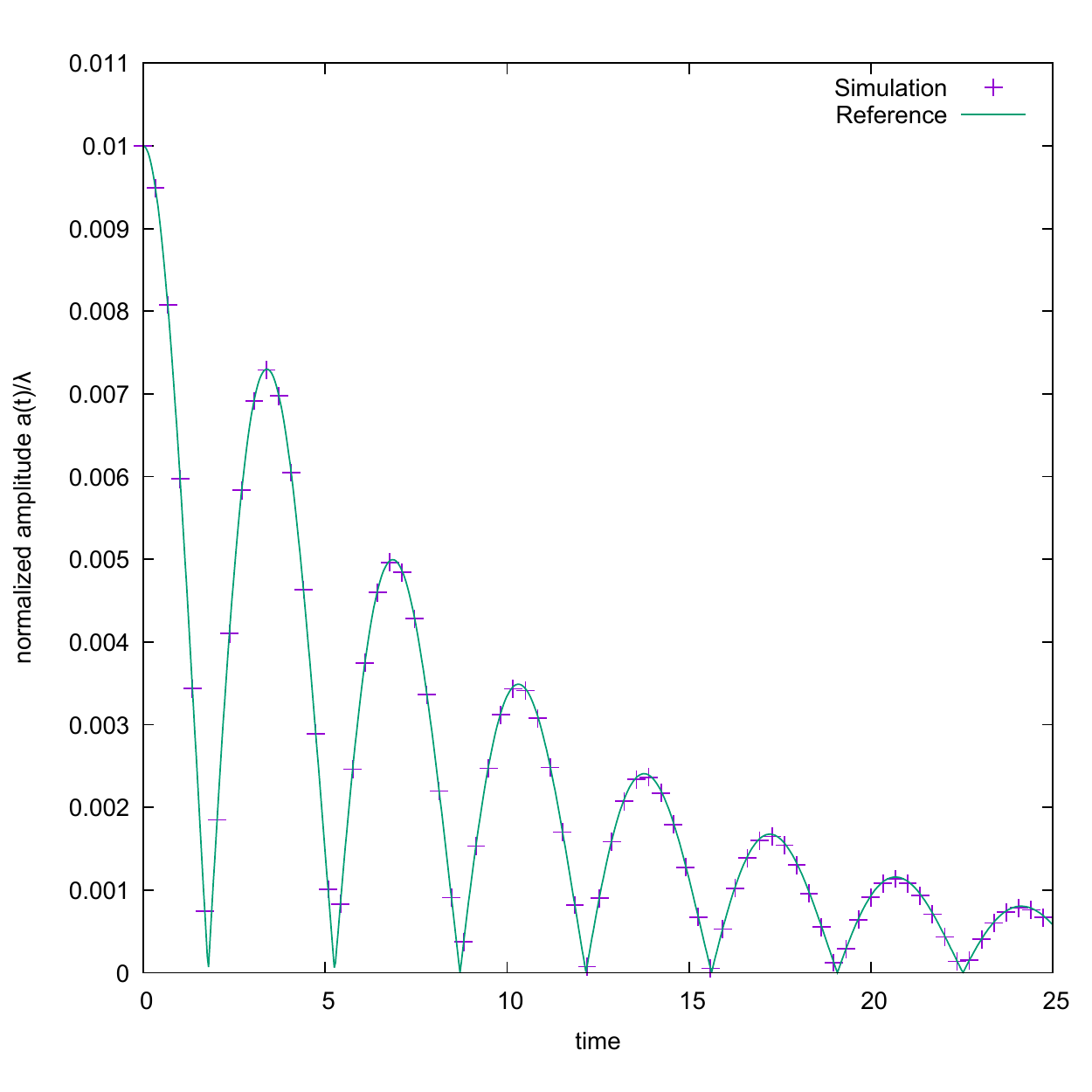}
\end{center}
\caption{Comparison of the temporal evolution of the \red {normalized} amplitude of 
the \red {capillary wave} obtained numerically and \red{ of} the analytical 
initial-value solution by \cite{prosperetti81}.}
\label{capwaves}
\end{figure}

In this section we present results of the oscillation of planar capillary
waves between two viscous fluids with equal density and viscosity in the presence 
of surface tension. The interface between the two fluids is slightly perturbed with 
a sinusoidal function of small amplitude $a_0$ and the initial velocity is set to 
zero. The solution of this problem is governed by the  Laplace number which is $La=\sigma\rho\lambda/\mu^2=3000$,
where $\lambda$ is the wavelength of the  sinusoidal function.
Simulation are performed in a box of dimensions $L_x=\lambda$ and $L_y$. 
The results are compared to the analytical initial-value solution (AIVS)
obtained in \cite{prosperetti81,denner2017dispersion} for small-amplitude capillary 
waves in viscous fluids.
In the AIVS the aspect ratio $L_y/L_x$ should be sufficiently large (at least 
\red {a value of} 2) 
and the initial capillary wave amplitude $a_0$ sufficiently small. Moreover the time 
step $\tau$ and the tolerance of the solvers should be at convergence. We have 
checked that all these four parameters were at convergence
for fixed $h$. We then look at the dependence of the remaining error on $h$. 
Figure \ref{capwaves} compares the temporal evolution of the amplitude of the 
interface perturbation with the AIVS.
\begin{table}
\begin{center}
\begin{tabular}{cccccc}
\hline
\hline
$N=\lambda/h$       &    8      & 16     & 32     & 64        & 128 \\
\hline
Gerris              &  0.159    & 0.032  & 0.0077 & 0.0022    & $5.5\,10^{-4}$\\
Basilisk            &  0.139  	& 0.024  & 0.0069 & 0.0024    & $4.8\,10^{-4}$  \\
Paris               &  0.050    &  0.023 & 0.0054 & 0.0015    & $4.1\,10^{-4}$ \\
\hline
\hline
\end{tabular}
\end{center}
\caption{Relative $L_2$ error of the numerical solution for capillary waves for various
codes and numbers of grid points $N$ per wavelength. The errors estimated by the codes
have been rounded to the nearest digit. Results for Gerris are from the website 
http://gfs.sf.net, not from the paper \cite{popinet09}. Results for Basilisk, 
\red {have been} obtained by the authors from the code published on the website 
{http://basilisk.fr}. 
}
\label{capwavesconv}
\end{table}

\begin{figure}
\begin{center}
\includegraphics[width=0.9\columnwidth,angle=0]{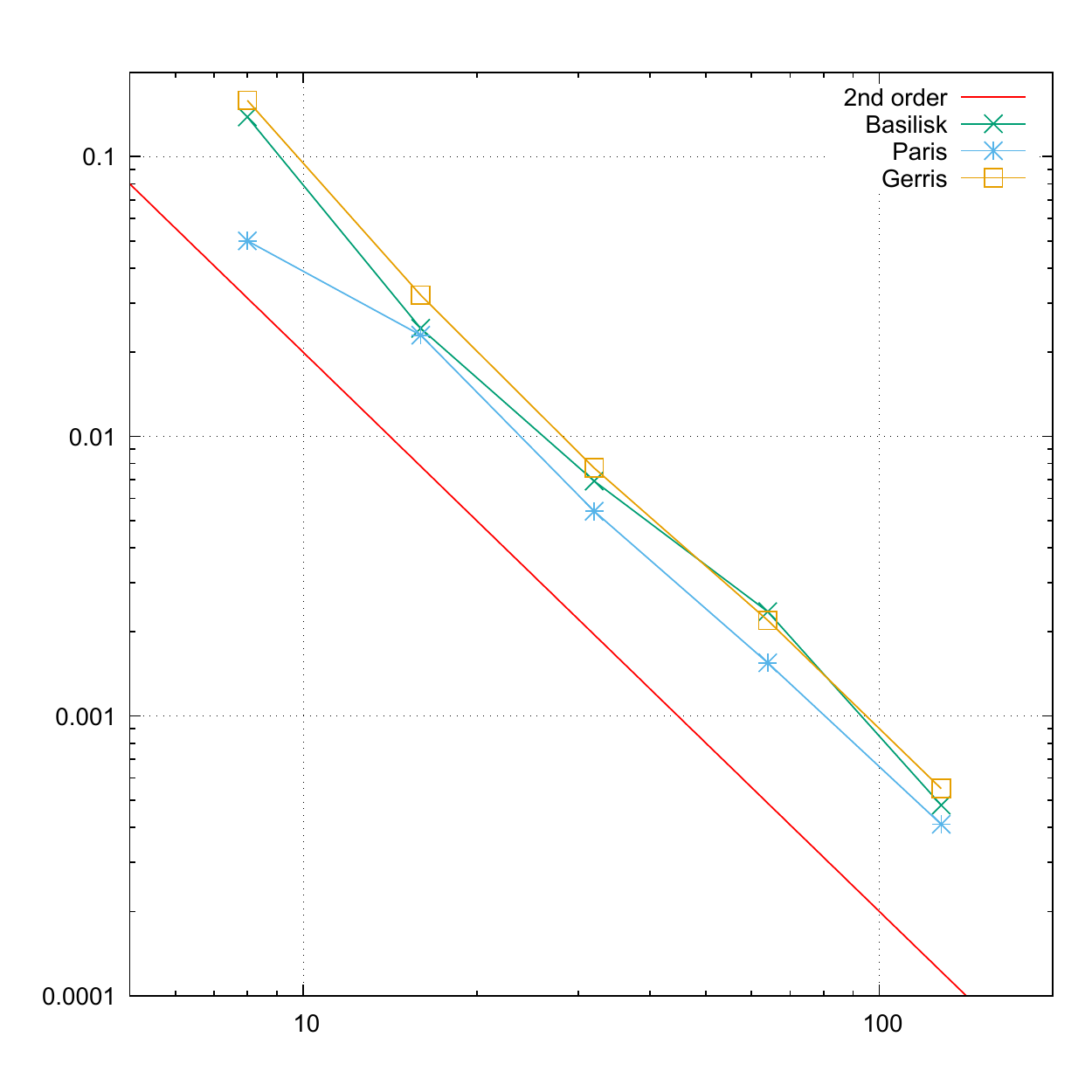}
\end{center}
\caption{Relative $L_2$ error of the numerical solution as a function of the number 
of grid points per wavelength $N=\lambda/h$ for the capillary wave \red {test}.}
\label{capwavesfig}
\end{figure}
The relative $L_2$ error norm of the difference between the numerical \red {solution
of a few numerical codes and the AIVS has also been} computed. 
The results, depicted in Table \ref{capwavesconv} and Fig. \ref{capwavesfig},
show second-order convergence for coarse grids. For very refined grids, when
the error is below 1\%, the accuracy on the solution is controlled by
the initial amplitude of the wave. In the case of the resolution $\lambda/h=128$,
instead of a $0.01$ amplitude as in 
\cite{popinet09}, a 0.005 amplitude had to be used to match the AIVS.

\subsection{Oscillating droplets and bubbles}
\begin{table}
\begin{center}
\begin{tabular}{cccccc}
\hline
\hline
 $D$ & $\mu_l$ & $\mu_g$ & $\rho_l$ & $\rho_g$ & $\sigma$  \\
$(m)$ & $(kg\,  m^{-1} s^{-1})$ & $(kg\, m^{-1} s^{-1})$  & $(kg\, m^{-3})$ & $(kg\, m^{-3})$ &  $(kg\, s^{-2})$ \\
\hline
$3\,\, 10^{-3}$ & $10^{-3}$ & $1.7 \, \,10^{-5}$& $10^{3}$ & $1.2$ & 0.0728 \\
\hline
\hline
\end{tabular}
\end{center}
\caption{Physical parameters (defined in the text) for the oscillating droplet.
}
\label{PhysicalParam}
\end{table}

\begin{table}
\begin{center}
\begin{tabular}{ccc}
\hline
\hline
  $r$ & $m$  & $\La$ \\
  $\rho_l/\rho_g$  & $\mu_l/\mu_g$ & $\sigma \rho_l d / \mu_l^2$ \\
\hline
833.3 & 58.82 & 218400 \\
\hline
\hline
\end{tabular}
\end{center}
\caption{Dimensionless parameters for the oscillating droplet,
$\La$ is the Laplace number.}
\label{dimensionlessParam}
\end{table}

\begin{figure}
\begin{center}
\includegraphics[width=0.9\columnwidth,angle=0]{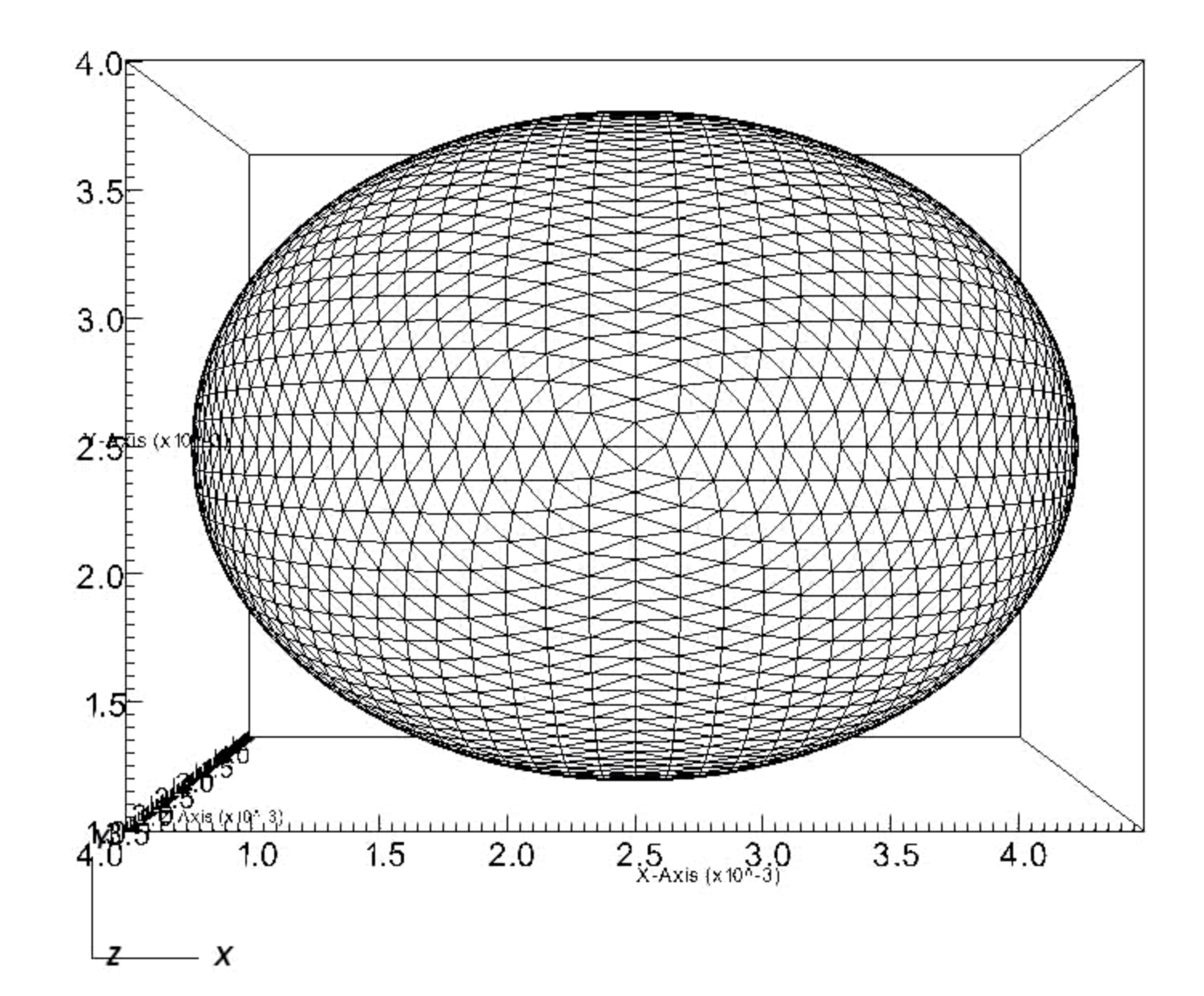}
\end{center}
\caption{Initial ellipsoidal shape of the droplet or bubble with the Front.}
\label{frontbubble}
\end{figure}

A droplet with a large density ratio is initialized with a small
ellipsoidal deformation. The droplet has air-water properties
described in Tables \ref{PhysicalParam} and
\ref{dimensionlessParam}. The initial shape, when tracked with the
Front, is shown in Fig. \ref{frontbubble}. This test is not designed
to assess the accuracy of the code, since the \red {implemented numerical} 
methods have already been used and tested elsewhere (see for example
\cite{Fuster:2009hg} for \red {an} oscillating droplet test with \red {the VOF
method} and \cite{Torres:2000hu,Olgac:2013fi} for similar tests
with Front Tracking. \red {However, it should be noted that in these}
references the tests are 2D, hence easier). We thus expect the
accuracy to be similar to that of already published and tested codes
using similar curvature and surface tension methods. The purpose of
the test is rather to ensure that the code is working as expected, and
to assess whether air-water properties, which are often creating
stability problems, are in fact in the stable regime of the code.

\begin{figure}
\begin{center}
\includegraphics[width=0.9\columnwidth,angle=0]{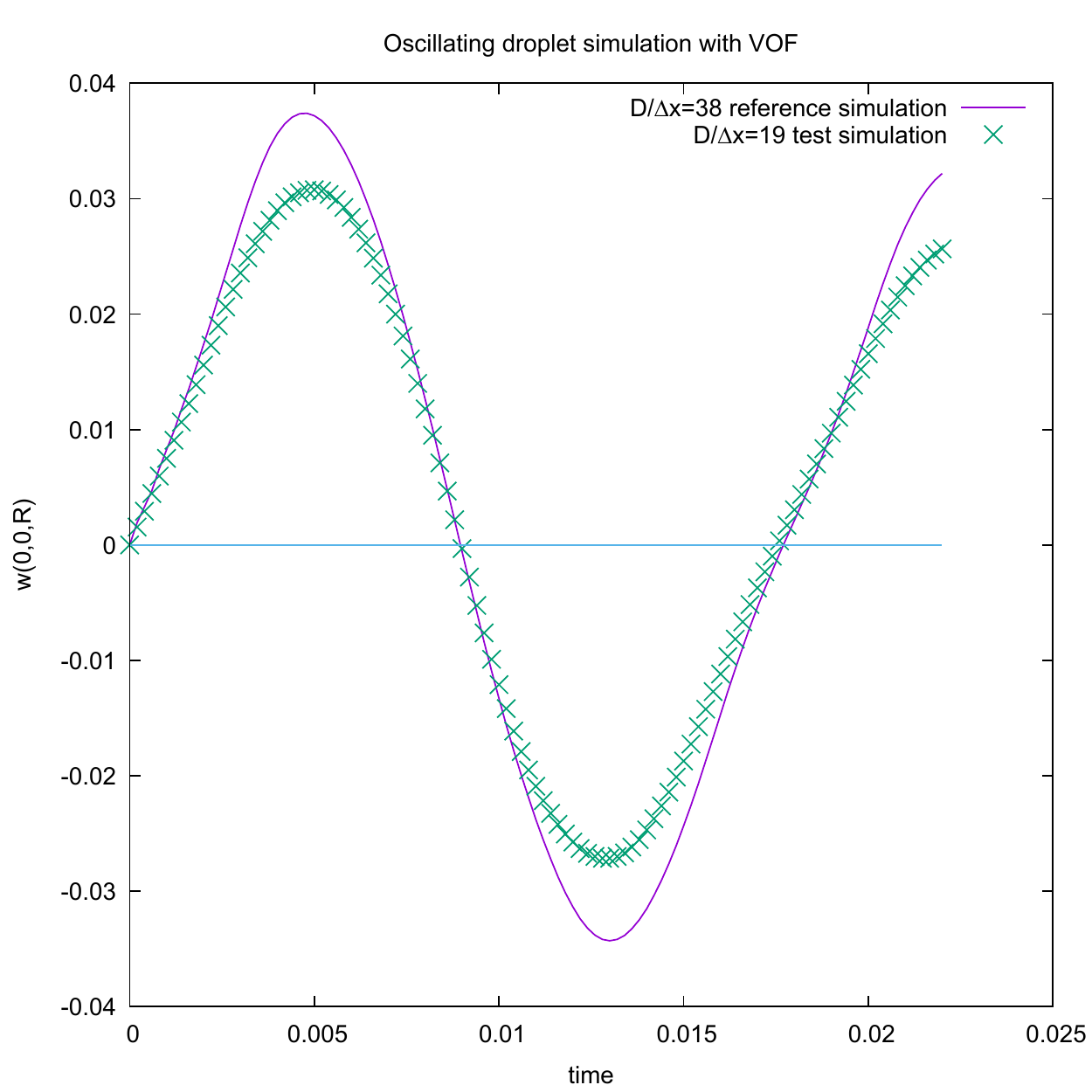}
\end{center}
\caption{Amplitude of capillary oscillations of the {\sf Droplet} test case.}
\label{oscidrop}
\end{figure}

\begin{figure}
\begin{center}
\includegraphics[width=0.9\columnwidth,angle=0]{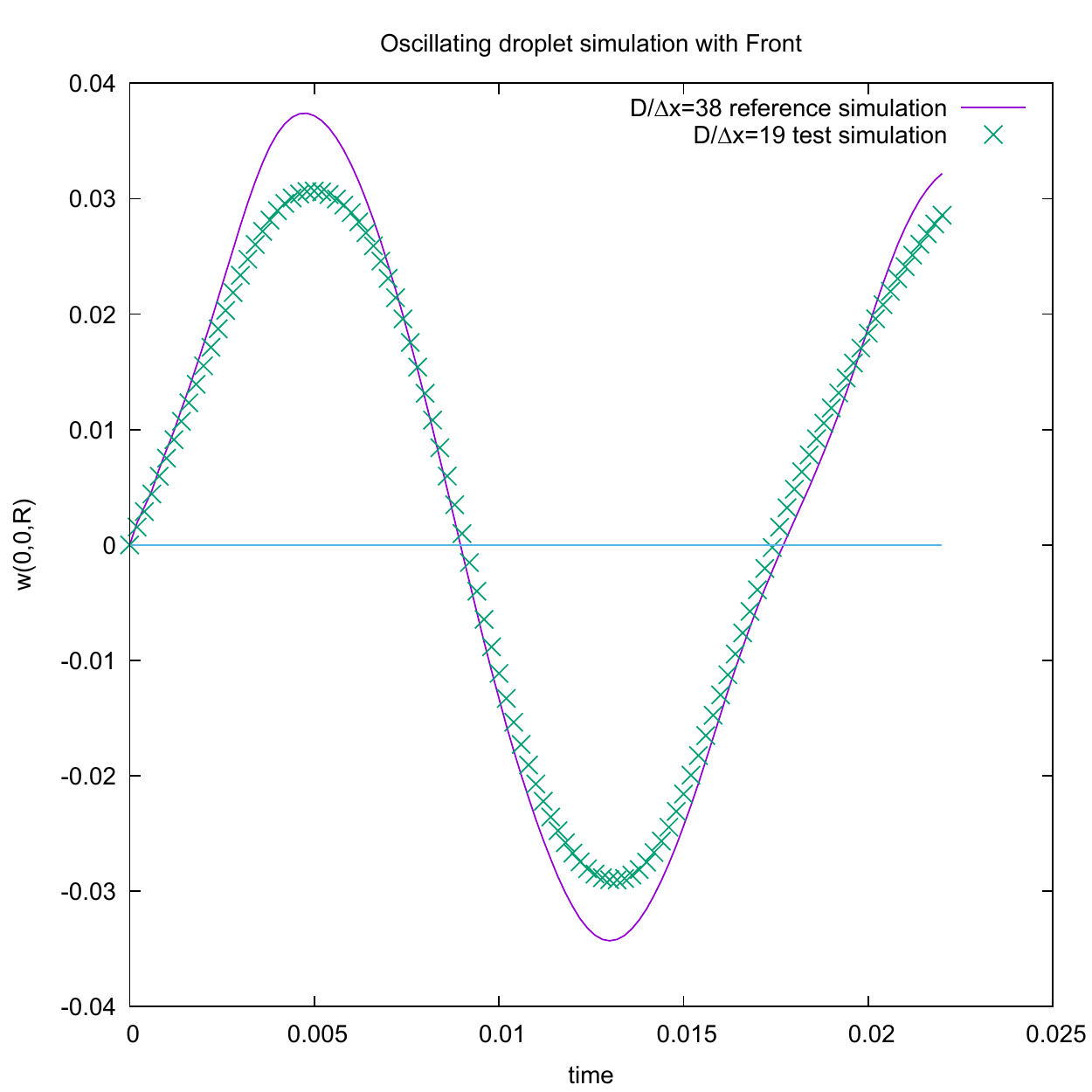}
\end{center}
\caption{Amplitude of capillary oscillations of the {\sf FrontDroplet} test case.}
\label{frontdroplet}
\end{figure}

Figure \ref{oscidrop} shows the amplitude oscillations \red {as a function of}
time for a droplet of $D/h=19.2$ grid points per diameter and an initial 
excentricity of $0.75$. The test is run without the \red {mass-momentum consistent} 
option, which is here seen to be unnecessary for the stability, and with the
mixed-height option discussed in  \ref{height-mixed}. The reference solution, 
plotted alongside the test simulation result, is obtained from the same VOF 
simulation at the larger resolution $D/h=38.4$. Satisfying agreement is found. 

Note that when this test is run automatically in the test suite, 
the reference solution stored in the {\sf Test/Droplet} directory has been obtained
by our code running in the same conditions, a device frequently used
in several test cases in the code test suite. That way, one tests that
the behavior of the code has not changed drastically, but not that the
code (original version and current version) is correct. 

We test the Front-Tracking part of the \red {code} by simulating the same droplet 
\red {test}. Results are shown on Fig. \ref{frontdroplet}. The reference solution
\red {is once again the VOF simulation at larger resolution ($D/h=38.4$).} 
Satisfying agreement is found \red {in this test as well}. 

We repeat these tests by just inverting the phases, thus initializing an air bubble
inside water. The physical and the numerical parameters (such as the scheme options) 
are  the same as before.

\begin{figure}
\begin{center}
\includegraphics[width=0.9\columnwidth,angle=0]{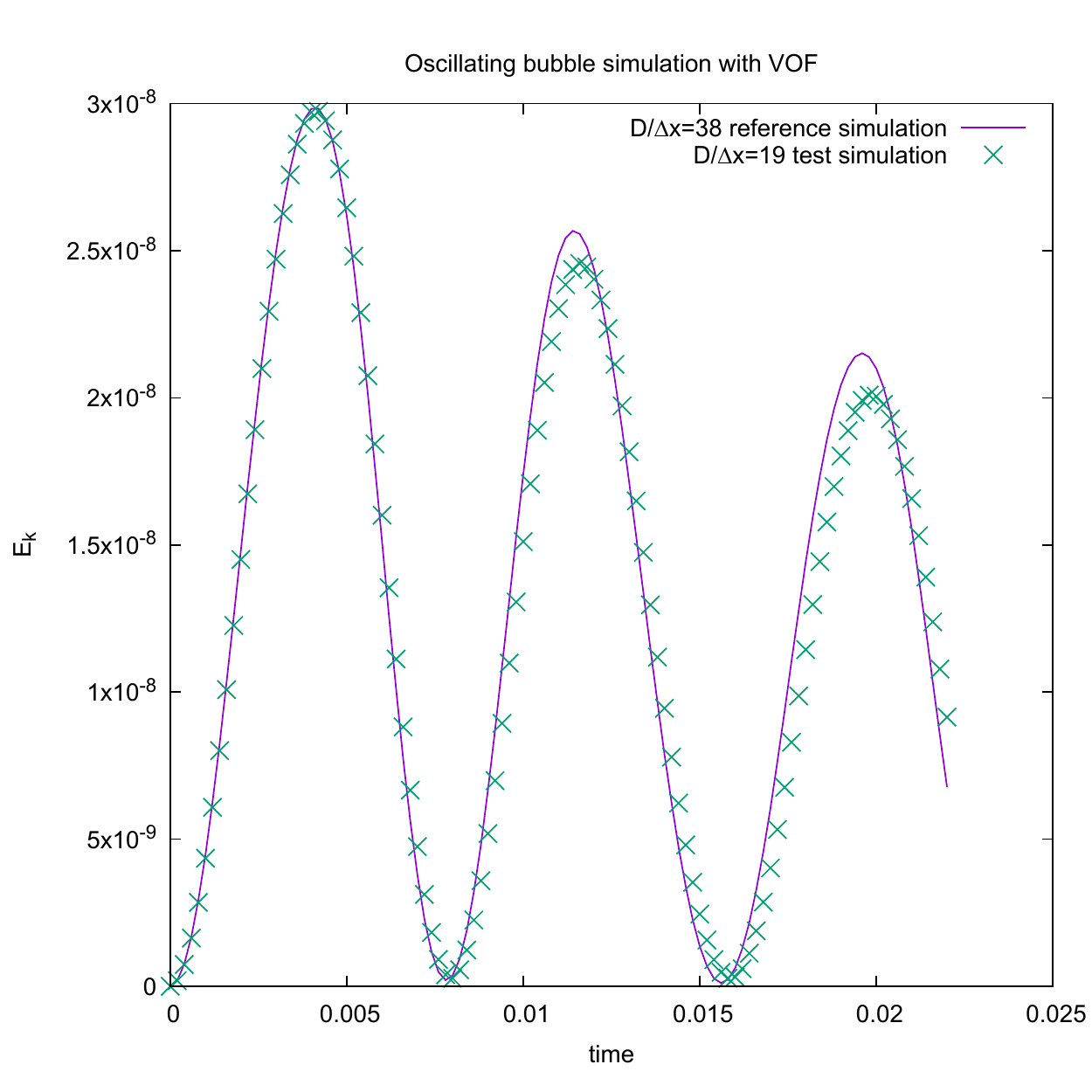}
\end{center}
\caption{Kinetic energy of  capillary oscillations of the {\sf Bubble} test case.}
\label{oscibubble}
\end{figure}

\begin{figure}
\begin{center}
\includegraphics[width=0.9\columnwidth,angle=0]{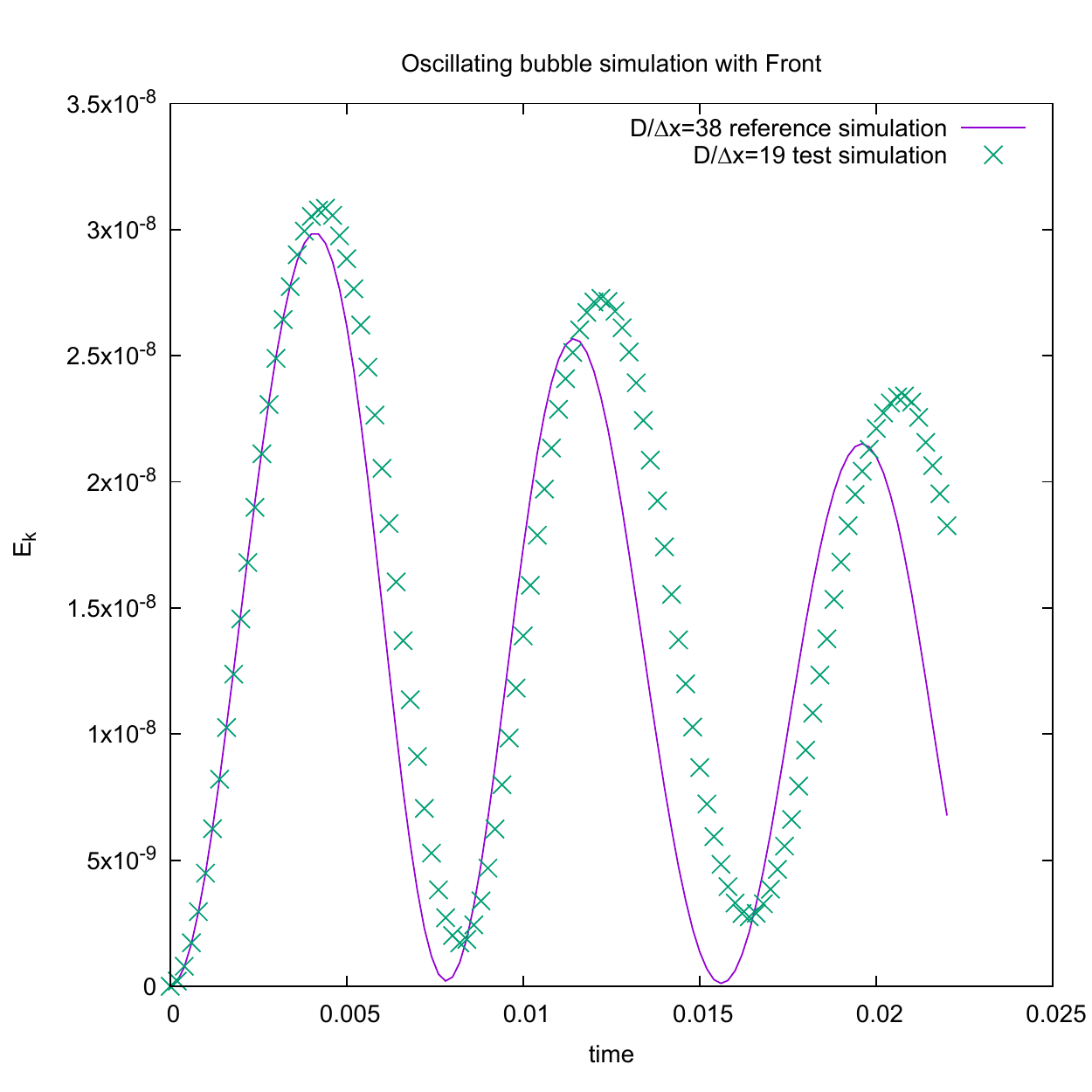}
\end{center}
\caption{Kinetic energy of capillary oscillations of the 
{\sf FrontBubble} test case.}
\label{frontbubbleosc}
\end{figure}

Figure \ref{oscibubble} shows \red {for the {\sf Bubble} test case
the kinetic energy oscillations as a function of} time for a bubble of $D/h=19.2$
grid points per diameter and an initial excentricity of $0.75$. 
The kinetic energy is used this time instead of the deformation amplitude.
The reference solution is again obtained from the VOF simulation at larger 
resolution, with $D/h=38.4 $ grid points. There is again good agreement. 

We test \red {again} the Front-Tracking part of the \red {code} by simulating the same 
bubble \red {test}. Results are shown on Fig. \ref{frontbubbleosc}. \red {The
reference solution is again that of the previous case (VOF simulation with resolution
$D/h=38.4 $)}. The agreement is satisfying, but a small drift of the kinetic energy 
is observed.

\subsection{Falling raindrop}
\begin{figure}
\begin{center}
\includegraphics[width=0.9\columnwidth,angle=0]{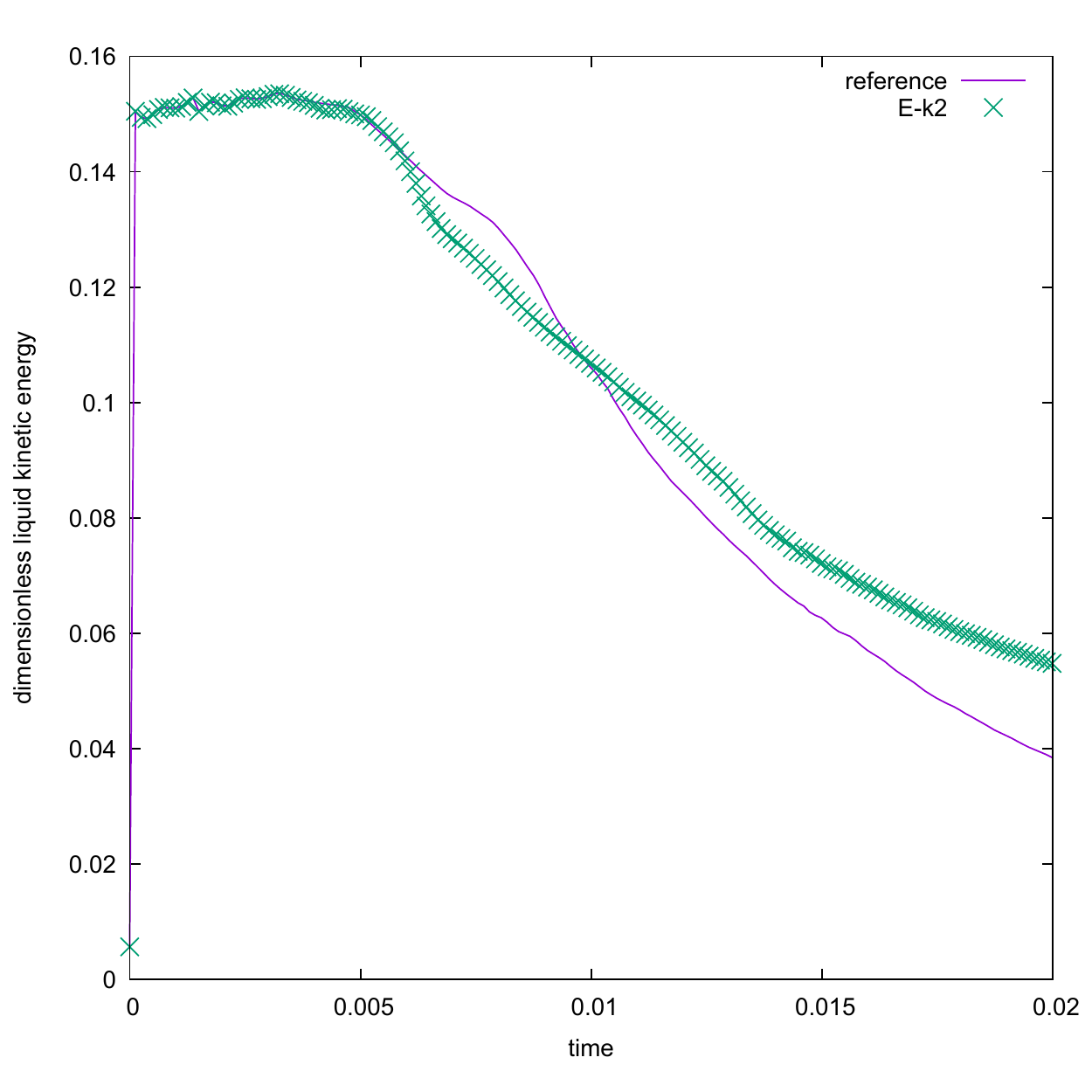}
\end{center}
\caption{Kinetic energy of a $3\,$mm falling raindrop at low resolution  
$D/\Delta x=8$.}
\label{raindrop}
\end{figure}

\begin{figure}
\begin{center}
\includegraphics[width=0.9\columnwidth,angle=0]{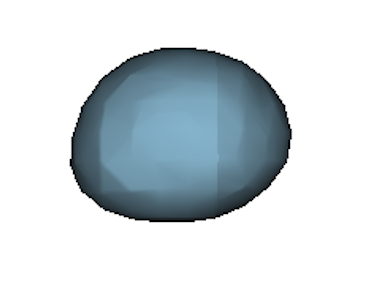}
\end{center}
\caption{Simulated shape of a $3\,$mm falling raindrop at low resolution  
$D/\Delta x=8$.}
\label{raindrop-shape}
\end{figure}

This is an important test case, since it is very demanding for several
codes. A spherical raindrop is setup with a diameter $d=3\,$mm and allowed
to fall in air under gravity. The droplet should remain approximately 
spherical, with \red {at most} a pancake or bun-like shape, but 
\red {for many numerical codes this case is rather difficult} and spurious
atomisation of the droplet is seen. For an example with the \basilisks 
code see \cite{pairettibag}. For details about the setup of the test we refer the 
reader to \cite{fuster2018momentum}. The test is performed with the parameters 
of Table \ref{PhysicalParam2}. Notice that these parameters differ slightly 
from those of \cite{fuster2018momentum} as we input here \red {the values of 
the physical parameters for air and water at temperature $T=20\,^\circ C$}.
The grid resolution is low $D/\Delta x=8$, since the test is more demanding 
(\red {i.e.,} it leads to a blowup of the code more easily) \red {at} low
resolutions. Figure \ref{raindrop} provides the result of the {\sf Raindrop} 
test in the code distribution. It is seen that the solution deviates somewhat 
from the reference,
however this is not 
worrisome,  as explained  in the introduction to this section \ref{testing}, since the flow
is in a regime that is very sensitive to 
\red {physical} parameters and initial conditions and any small change in the 
code will create a deviation of \red {that} sort. The shape of the droplet is 
shown in Fig. \ref{raindrop-shape}. 
\begin{table}
\begin{center}
\begin{tabular}{cccc}
\hline
\hline
$\mu_l$ & $\mu_g$ & $\rho_l$ & $\rho_g$   \\
$(kg\,  m^{-1} s^{-1})$ & $(kg\, m^{-1} s^{-1})$  & $(kg\, m^{-3})$ & $(kg\, m^{-3})$  \\
\hline
$1.0016 \,10^{-3}$ & $1.835 \, \,10^{-5}$& $998.2$ & $1.19$ \\
\hline
\hline
\end{tabular}
\end{center}
\caption{Physical parameters for the raindrop test. Only the parameters that differ from
  Table \ref{PhysicalParam} are given. 
}
\label{PhysicalParam2}
\end{table}

\section{Installation and Usage} 

\subsection{External Libraries}
When solving elliptic equations, we may apply the {\sc Hypre} package SMG 
\cite{hypre02}, a semi-coarsening multigrid solver with 3D plane smoothing on structured,
cuboid meshes, as mentioned in Section \ref{hls}.

Installation of the static library versions (\textsf{*.a} files) of {\sc Hypre} is 
controlled via the Makefile. For example, for a Linux system, this can be realized 
via the line
\begin{small}\begin{verbatim}HYPRE_DIR = $(HOME)/some_path/hypre-2.10.0b/src/lib \end{verbatim} \end{small}
in the Makefile. Note that the file \texttt{libHYPRE.a} is placed in the directory \texttt{HYPRE\_DIR}. Consequently, the actual linking is ensured by another Makefile 
line
\begin{small} 
\begin{verbatim}HYPRE_LIBS = -L$(HYPRE_DIR) -lHYPRE\end{verbatim}
\end{small}
which is specified in the default \pariss distribution. Note that the {\sc Hypre} 
version used for the majority of \pariss development has been 2.10.1; it has been tested
for stability in serial and massively parallel runs \cite{anisz18}. In case of 
{\sc Hypre}-related problems, fallback to version 2.10 is recommended for debugging.  

The {\sc Vofi} library \cite{bna2015numerical,bna2016vofi}  \red {may be used to initialize
the volume fraction field $C$ in 2D and in 3D} (see Sec. \ref{vofi}). It is an interesting
option in cases where initial conditions depend strongly on a very precise interface 
geometry, e.g. free-surface solutions of bubble dynamics \cite{malan19, anisz18}, and \red
{more rarely when initialization requires a considerable amount of computational time 
compared to Paris simulation time} (as in the curvature test case). Linking of this library 
is performed in the exact same fashion as in the case of {\sc Hypre}. The static library
file is named  \texttt{libvofi.a}, and the respective linker switch is \texttt{-lvofi}.

\red {Since} {\sc Hypre} and {\sc Vofi} are \red {indeed} optional, compilation 
without them is made easy. The user may set or unset the variables {\sf HAVE\_VOFI} and 
{\sf HAVE\_HYPRE} in his shell prior to compiling. If these variables are set the 
Makefile will attempt to locate the corresponding libraries, \red {if they are not set
the compilation proceeds} without the libraries. The fallback for {\sc Hypre} is the 
built-in Gauss-Seidel solver followed by the in-code multigrid solver, and the fallback 
for {\sc Vofi} are the built-in VOF initialization procedures.

\subsection{Output Files} 
Various output formats are available in the code: VTK, SILO, and MPI I/O. While the VTK
option generates ASCII files, the latter two produce binary files. The output in SILO 
format is done based on the SILO library developed \red {at} LLNL. For both the VTK and SILO 
output options, the independent parallel approach is used, namely every MPI process 
generates a separate file. This will become an issue for large-scale simulations using 
\red {many} MPI processes, since creating a large number of small files simultaneously 
may crash the indexing server. 
An output option based on the MPI I/O library is implemented in PARIS for large-scale
simulations, which adopts, instead, a cooperative parallel approach and creates a single 
file for each variable for each output. A post-processing code was developed in PARIS to
convert the MPI I/O outputs and SILO files offline for visualization. This code is 
available in the distribution as the file {\sf util/post\_utility.f90}.

\subsection{Input Files} 
\pariss requires a set of input files (in text format) to initialize and start the 
simulation. These files are
\begin{itemize} 
\item \textsf{input} - global and front-solver parameters.
\item \textsf{inputFS} - free-surface solver parameters (optional, free-surface 
simulations only). 
\item \textsf{inputlpp} - Lagrangian \red {point-particle} module parameters (optional, implicitly requires two-phase flow, see \cite{ling15}). 
\item \textsf{inputvof} (optional, VOF parameters, as above).
\item \textsf{inputsolids} (optional, solid objects parameters).
\end{itemize}
All input files contain over 220 parameters, thus listing all of them is not practical 
in this paper; instead we will only point to general rules governing the use of these 
files. \red {The} reader may find the default values of these parameter in the source code, 
usually just below the \textsf{namelist} instruction.

All the input lines contain the parameter specifications written as ``variable = value'', 
with values being reals, integers, string  or boolean (T/F). In most cases (in the code
versions distributed in main darcs tree) the variables are commented Fortran-style, 
i.e. in the same line, following an exclamation mark, for example

\begin{small}
\begin{verbatim}
MaxDt   = 5.e-5  ! maximum size of time step 
dtFlag  = 2      ! 1: fixed dt; 2: fixed CFL
dt      = 1.0e-4 ! dt in case of dtFlag=1
MAXERROR= 1.0d-6 ! Residual for Poisson solver
CFL     = 0.042	   
\end{verbatim}
\end{small}

It must be noted that the \pariss source code uses the Fortran \textsf{namelist} input 
type, consequences of that being that all input files have ``free format'', i.e. lines 
can change order or be deleted. All variables are initialized to default values in the
source code. Thus, \pariss will initialize with an empty input file, however in such a 
case the simulation will be short. Indeed by default \textsf{Nx=0} (the grid has zero 
points in $x$ direction) in order to prevent simulations with some of the absurd input 
files that could be selected by mistake. Thus a minimum input file should contain at 
least a specification of \textsf{Nx}. For more demanding simulations, beginner users 
are encouraged to familiarize themselves with input file examples, such as templates 
found in the \texttt{Tests} sub-directory in the distribution which can be copied and 
modified to create new \pariss cases.
Note that in the Test suite, input files are often generated from template files 
such as \texttt{inputfilename.template} using shell scripts.

\section{Acknowledgements} 

We thank  Dr. V. Le Chenadec, Mr. C. Pairetti, Dr. S. Popinet and Dr. S. Vincent for useful
conversations on the topics of this paper.  

Portions of this  work were supported by National Science Foundation Grants CBET-1335913 and NSF DMS-1620158, 
by the ANR MODEMI project (ANR-11-MONU-0011) program and by grant SU-17-R-PER-26-MULTIBRANCH
from Sorbonne Universit\'e. This work was granted access to the HPC resources of TGCC-CURIE,
TGCC-IRENE and CINES-Occigen under the allocations  t20152b7325,  t20162b7760, 2017tgcc0080
and A0032B07760, made by GENCI and TGCC. 
The authors would also like to acknowledge the MESU computing facilities of Sorbonne
Universit\'e. Finally, the simulation data are visualized by the software VisIt developed 
\red {at} the Lawrence Livermore National Laboratory.




\bibliographystyle{elsarticle-num} 
\bibliography{multiphase}

\appendix
\section{Details of curvature computation from \red {height functions (HF)}} 
\label{app:A}

\subsection{Height computation}

The height computation is performed as described in Section \ref{height_function}.
A more general definition than (\ref{hfeq1}) is to consider for each cell 
$\Omega_{i,j,k}$ the possible existence of one of six height functions defined with 
reference to a direction $\E_a$, $1\le a\le 3$, where $\E_a$ is one of the 
\red {Cartesian} base vectors \red {aligned} with the grid, and an orientation
\red {$\eps = - {\rm sign} (\E_a \cdot \nabla C)$} (the minus sign ensures that the 
canonical situation where the \red {``liquid''} $C=1$ is below the ``air'' $C=0$ has 
$\eps=1$. It also corresponds to the sign convention for the interface normal 
$\N\delta_S = - \nabla \Heaviside$). This cell-and-orientation-dependent HF is defined 
as
\be
H_{i,j,k}^{(a,+)} = \sum_{{\rm stencil}\, S} C_{l,m,n} - L_{i,j,k} 
\label{hfeq2}
\nd
where the sum is over all the cells in a \red {one-dimensional symmetric} ``stencil'' 
or ``stack'' $S$ centered on $\X_{i,j,k}$ and oriented parallel to $\E_a$ and for the 
``positive'' orientation $\epsilon$. 
\red {$L_{i,j,k} = \eps (\X_{i,j,k} - \X_O)\cdot \E_a$ is the distance} from the base 
of the stack to the cell center $\X_{i,j,k}$. When the orientation is $\eps=-1$ 
\be
H_{i,j,k}^{(a,-)} = \sum_{{\rm stencil}\, S} \left( 1 - C_{l,m,n}\right) - L_{i,j,k}
\nd
and  the distance $L_{i,j,k}$ is now with reference to an origin in direction $- \E_a$ 
from the cell. An example of stack is shown on Fig. \ref{floatsam}(c). 
This height function can be computed whenever \red {both} the bottom cell and the top 
cell of the stack do not contain the interface, and the interface crosses only once 
the intermediate cells. This can be tested by requiring that there are cells with 
$C=0$ and $C=1$ at the top and bottom \red {of the stack (see again 
Fig. \ref{floatsam}(c)).}

For a straight line interface the height function is  exact. It is interesting to see 
how many cells are needed in the stack $S$ to find the height for a straight line in 2D. 
The most ``difficult'' case is the \red {$45^\circ$} case. Thus, considering a cell 
crossed by 
the interface, one \red {should} explore one cell above \red {and} one cell below that 
cell. With the addition of the full and empty cells this requires the {\em exploration} 
of two cells above and below the starting cell. The total number of cells for a symmetric 
stencil about the starting cell would thus be five, but the total number of cells in a 
stencil maybe as low as \red {three}. However, \red {even with} a vanishing amount 
of curvature five cells in a symmetric stencil are not sufficient and seven cells are 
needed. Thus one would need to explore $N_d = 3$ cells above \red {and} below. 
In three dimensions the most ``dangerous'' cases now have the plane normal $\N=(1,1,1)$. 
A similar reasoning also leads to a height of seven cells for the symmetric stencil in 3D. 
Note that the stencil does not have to be symmetric, rather this is a accidental feature 
or our implemention of the method.

For each cell, it is determined whether there is a full or empty cell at a maximum distance 
$N_d$ (the parameter {\sf NDEPTH} in the code) above or below the cell. The value 
{\sf NDEPTH=3} is \red {hard-coded}. In order to function in parallel, and since only two 
layers of \red {cells} are exchanged between MPI processes, the sum in (\ref{hfeq1}) is 
broken in two parts, one in each processor. Then the processes exchange two \red {pieces
of information}, the ``partial sums'' \red {just} computed and the lengths $L_{i,j,k}$ in 
expression (\ref{hfeq2}), allowing them to reconstruct the full sum. 

\subsection{First pass, first attempt: \red {fully-aligned} heights}

In the first pass, a loop is performed over all cells \red {$\Omega_{i,j,k}$} cut by the 
interface, hence having $0 < \cijk < 1$. In \red {the} first pass two attempts are made. 
\red {Let the current cell be $\Omega_0$ with} grid coordinates $i_0,j_0,k_0$. 
In the first attempt, the normal $\N$ is estimated by MYCS \red {\cite{aulisa07}}. Then 
the grid direction $\E_a$ closest to the normal is determined (maximum of 
\red {$|\N \cdot \E_a|$}). Without loss of generality \red {we}
consider the case $a=3$ and the \red {horizontal plane perpendicular to $\E_3$,
that is} the grid plane most \red {closely} aligned with the interface.
A $3\times 3$ planar \red {block} of cells $\Sigma_0$, aligned with this plane, 
is selected containing \red {the nine} cells $\Omega_{i,j,k_0}$, \red {with} 
$i_0-1 \le i \le i_0+1$ and $j_0-1 \le j \le j_0+1$. For all these cells, either a 
height $H_{i,j}$ is readily available, or is searched in the above and below cells over 
two layers, that is for $k_0 \pm 1$ and $k_0 \pm 2$. When all nine heights are available, 
the \red {following} coefficients of the polynomial (\ref{apoly}) can be found using 
\red {finite differences}
\begin{align}
  a_1 &= \partial^2_{xx} H \simeq H_{1,0}-2H_{0,0}+H_{-1,0}\, , \nonumber \\
  a_2 &= \partial^2_{yy} H \simeq H_{0,1}-2H_{0,0}+H_{0,-1}\, , \nonumber \\ 
  a_3 &= \partial^2_{xy} H \simeq  (H_{1,1}-H_{-1,1}-H_{1,-1}+H_{-1,-1})/4\, , \nonumber \\ 
  a_4 &= \partial_x H \simeq (H_{1,0}-H_{-1,0})/2\, , \nonumber\\ 
  a_5 &= \partial_y H \simeq (H_{0,1}-H_{0,-1})/2\, .
  \label{apolycoef}
\end{align}

\subsection{First pass, second attempt: mixed heights}
\label{height-mixed}
If the first attempt fails then a ``mixed heights'' approach is used, but only if the 
parameter {\sf MIXED\_HEIGHTS} is set to {\sf 'T'}.  For every height 
$H_{i,j,k}^{(a,\eps)}$ that has been calculated, a point 
$\X_{i,j,k}^a = H_{i,j,k}^{(a,\eps)} \E_a + x_b \E_b + x_c \E_c$ is defined, where
$x_b$ and $x_c$ are the cell-central coordinates in the two directions other than $\E_a$. 
Since there are six height directions, up to 54 points could be computed. However, a 
general orientation is computed using the MYCS normal and only the orientations compatible 
with that orientation are retained, which yields 27 possible points. With certain point
configurations there is a risk of a degenerate case where the least-square linear operator 
is not invertible. This happens \red {for example} in the set of six points obtained with
combinations of $x=0,1$, $y=-1,0,1$, $z=0$. All paraboloids of the form $z=x(1-x)$ pass 
through these points. Other degeneracies are possible: points all on a circle will
be fitted by infinitely many revolution paraboloids $z'=\kappa(x^2+y^2)/2$. To avoid these
degeneracies, and after trial and error, the minimum number of points requested is hard-coded 
as $N_s=${\sf NFOUND\_MIN +1}$=7$. In addition to these ``mixed heights'' points, the 
centroid of the VOF face in the current cell $\Omega_{i,j,k_0}$ is added to the set of 
points \red {to be fitted}.  In some cases, different directions $\E_a$ could yield two 
close points, \red {say $\X_{i,j,k}^a$ and $\X_{l,m,n}^b$,} in the same cell or in a 
neighboring cell. \red {In this case,} if $||\X_{i,j,k}^a - \X_{l,m,n}^b||< h/2$ then 
one of the two points is rejected. Which point is rejected depends on the
order in which points are added to the stack, which in turn depends on the order in which 
mixed heights are investigated, typically closest to the general orientation. Before the 
fitting is performed, an approximate normal is computed using the MYCS approach 
and the coordinate system is rotated so that the $z$ axis is now aligned with the 
approximate normal. The rotation is optional and is controlled by the parameter 
{\sf DO\_ROTATION}. We \red {have found} that performing rotation \red {has} a certain 
positive influence on the accuracy of the results, although it is not clear why. 

\begin{figure}[!htb]
\begin{center}
\includegraphics[width=0.8\columnwidth,angle=0]{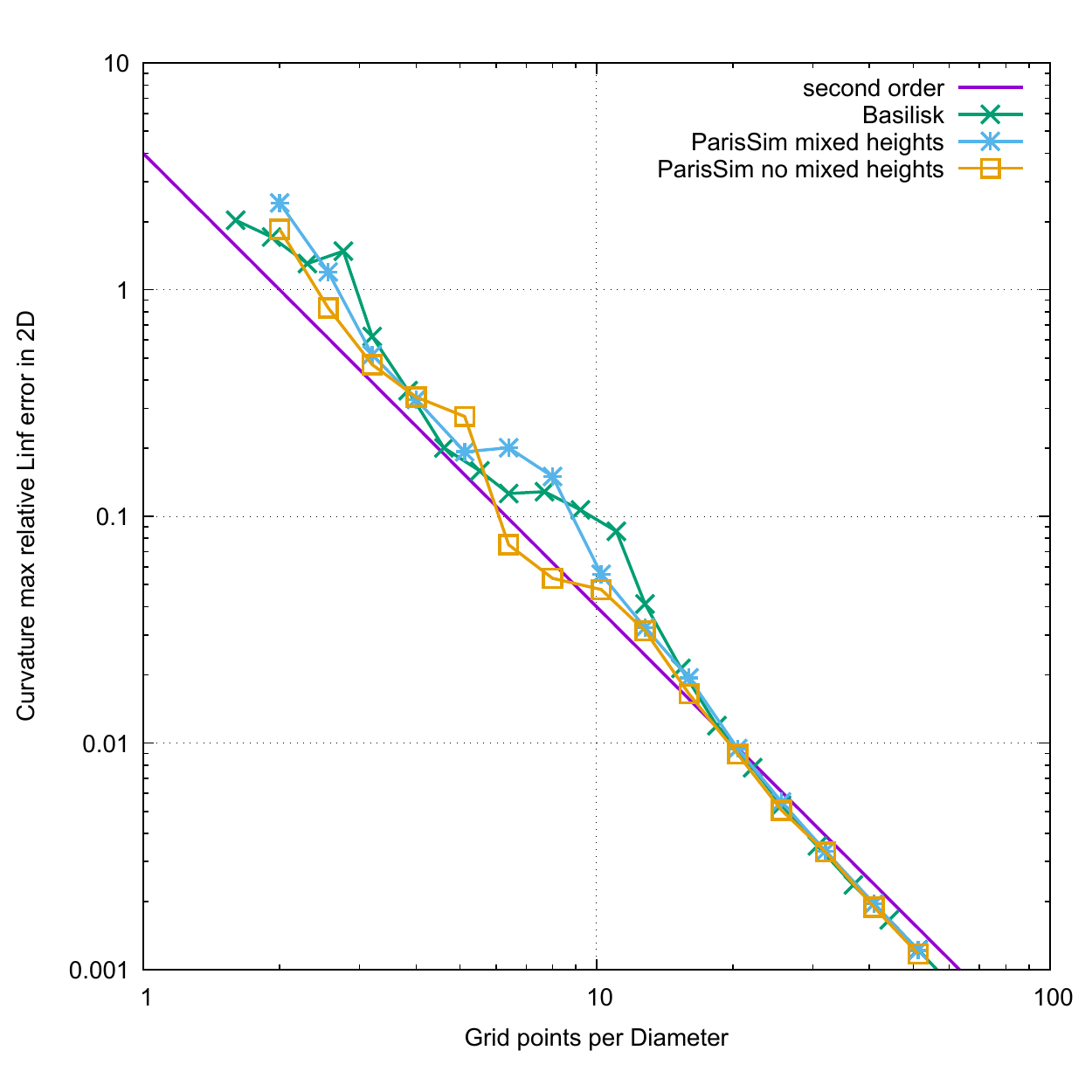}
\end{center}
\caption{Maximum $L_\infty$ error norm in two dimensions for the curvature estimated 
for a cylinder using the height function method in \pariss and \red {\basilisk}. 
Two \pariss results are shown, one with the mixed curvature option and one without. 
Averaging is used in both cases. Using the mixed-cell option yields less accurate 
results than not using it.}
\label{owkes2-mixed}
\end{figure}

By default the parameter {\sf MIXED\_HEIGHTS} is set to {\sf 'T'} (true).  This gives 
less accurate results in $L_1$ norm for curvature, but a smoother computation and as a 
result simulations appear to be more stable for large density ratios when the 
\red {mass-momentum consistent} scheme is not used. The results in Fig. \ref{owkes2} are 
with {\sf MIXED\_HEIGHTS}$=${\sf 'F'} . With {\sf MIXED\_HEIGHTS}$=${\sf 'T'} one obtains 
the results of Fig. \ref{owkes2-mixed}.  The results without mixed heights and those 
from \basilisks are added for comparison. There is a difference remaining with the 
\basilisks computations than we have not yet been able to explain.

\begin{figure}[htb]
\begin{center}
\includegraphics[width=0.8\columnwidth,angle=0]{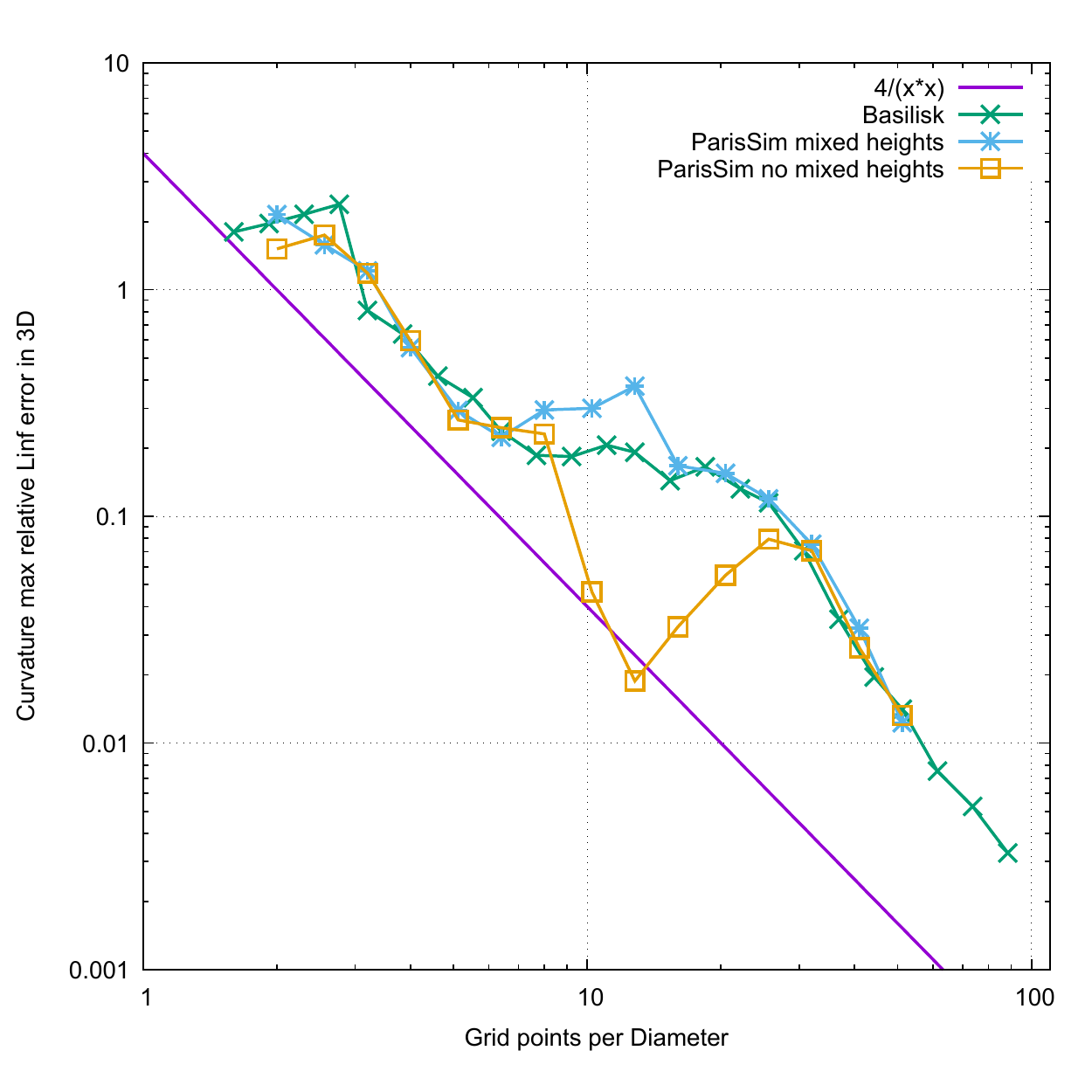}
\end{center}
\caption{Maximum $L_\infty$ error norm in three dimensions for the curvature estimated 
for a sphere using the height function method in \pariss and \basilisk. 
Two \pariss results are shown, one with the mixed curvature option and one without. 
Averaging is used in both cases. Using the mixed-cell option yields less accurate 
results than not using it.
}
\label{owkes3-mixed}
\end{figure}

\subsection{Averaging scheme}
\label{app:av}
A new loop over all cells cut by the interface is started. If both schemes above have 
failed in the current cell, an average is performed over neighboring cells that have
been \red {successful} by either method in the first pass. For each cell 
$\Omega_{i,j,k}$, the cubic set of neighbors
$$
B_{i,j,k} = \{\Omega_{l,m,n} \vert i-1 \le l \le i+1,j-1 \le m \le j+1, 
k-1 \le n \le k+1 \}
$$
is defined. If at least one of the cells in $B_{i,j,k}$ has been \red {successful}, 
the resulting curvature in  $\Omega_{i,j,k}$ is the average curvature of these 
\red {successful} cells.

\subsection{Second pass: centroid fit}

A final loop on cells $\Omega_{i,j,k}$ is performed. If all \red {previous} approaches 
have failed or are not set to be used, then one falls back to a fitting of centroids. 
In each cell of \red {the cubic set} $B_{i,j,k}$ containing an interface, the cell
centroid is computed using the {\sf vof\_functions} microlibrary included in the code (implementing the method in \cite{Scardovelli00}). Except for very small fragments, the 
interface must find at least five neighboring cells in addition to the current cell. 
This gives six points with which to fit the parameters in expression (\ref{apoly}).

In some cases, the fit fails because the least-square linear operator is not invertible. 
The code then reports in a statistical manner these failures and flags the cell as having 
an uncomputable curvature. A zero surface tension force is then added to the momentum.

\subsection{Comparison with other implementations of height-function curvature}

The accuracy contrast when modifying the mixed-heights option is even more dramatic in 
3D, see Fig. \ref{owkes3-mixed}. There is a striking drop in the $L_1$ error near 13 grid 
points per diameter. This drop is due to the averaging step in \ref{app:av}. Suppressing 
the averaging reverts the results to accuracies comparable to those of \basilisks or 
slightly better.

\end{document}